\documentclass[prb,twocolumn,superscriptaddress,a4paper,twoside]{revtex4}

\usepackage{amsmath}
\usepackage{amssymb}
\usepackage{todonotes}
\usepackage{nicefrac}
\usepackage{graphics}
\usepackage{color}
\usepackage{graphicx}
\usepackage{verbatim}
\usepackage{float}

\makeatletter
\newcommand*{\rom}
[1]{\expandafter\@slowromancap\romannumeral #1@}
\makeatother

\begin{document}

\title{Time-dependent exchange-correlation hole and 
potential of the electron gas}

\author{K. Karlsson}
\affiliation{Department of Engineering Sciences, University of Sk\"{o}vde, SE-541 28
Sk\"{o}vde, Sweden}
\author{F. Aryasetiawan}
\affiliation{
Department of Physics, Division of Mathematical Physics, 
Lund University, Professorsgatan 1, 223 63, Lund, Sweden}

\begin{abstract}
The exchange-correlation hole and potential of the homogeneous electron gas have
been investigated within the random-phase approximation, employing the plasmon-pole
approximation for the linear density response function.
The angular dependence as well as the time dependence of 
the exchange-correlation hole are illustrated for a Wigner-Seitz radius $r_s=4$ 
(atomic unit).
It is found that there is a substantial cancellation between exchange and
correlation potentials in space and time, analogous to the cancellation of
exchange and correlation self-energies. Analysis of the sum rule explains why
it is more advantageous to use a non-interacting Green function than a renormalized one
when calculating the response function within the random-phase approximation and 
consequently the self-energy within the well-established $GW$ approximation.
The present study provides a starting point for more accurate and comprehensive 
calculations of the exchange-correlation hole and potential of the electron gas
with the aim of constructing a model based on the local density approximation
as in density functional theory.
\end{abstract}

\maketitle

\section{Introduction}

The homogeneous electron gas has been a long-lasting and an invaluable 
model of valence electrons in solids.
Pseudopotential theory \cite{martin} explains why the behavior of valence electrons in solids, 
despite the presence of a strong ionic potential, nevertheless resembles that of the
electron gas.
The effects of exchange and correlations of the interacting homogeneous electron gas 
have been studied thoroughly for the last six decades \cite{hedin1965,mahan,martin-reining-ceperley}. 
An important empirical
observation is that the main features of the spectral
function arising from exchange and correlations are quite robust and can be carried over to
real materials. For example, setting aside strongly correlated systems \cite{aryasetiawan-nilsson},
it is generally the case that the spectral function of most materials 
consists of a quasiparticle peak and an incoherent satellite feature which can be 
traced back to collective charge excitations (plasmons), just as found in the electron gas
\cite{aryasetiawan1998}.

The effects of exchange and correlations have been traditionally studied using the concept 
of self-energy, a nonlocal and energy-dependent quantity that acts on the Green function
as a convolution in space and time \cite{fetter-walecka,negele}. 
In a recent development, a different framework for 
representing exchange and correlations was proposed in the form of a time-dependent
exchange-correlation (xc) potential \cite{aryasetiawan2022a,aryasetiawan2022b}. 
This formalism is fundamentally different from the
self-energy approach in that the potential acts locally or multiplicatively 
on the Green function. Most importantly, the potential arises 
naturally as a Coulomb potential of 
a charge distribution (exchange-correlation hole) which fulfills a sum rule and 
some exact properties. Moreover, due to the special form of the Coulomb interaction, 
which depends solely on the separation of two point charges, it can be shown that the
exchange-correlation potential is in fact the first radial moment of the spherical 
average of the exchange-correlation hole.
This appealing result
is the analog of the result found in the exact expression for the ground-state
exchange-correlation energy, 
which is very much utilized in density functional theory and partially explains
the success of the local density approximation \cite{gunnarsson1976,jones1989}.  

The exchange-correlation potential formalism also provides a simple physical picture of the 
propagation of an added hole or electron in a many-electron system as in photoemission and
inverse photoemission experiments. The added hole or 
electron induces a temporal density fluctuation of the system initially in its ground state,
giving rise to the exchange-correlation potential, which in turns acts on the Green function
representing the propagation of the added hole or electron.

In this paper, the time-dependent exchange-correlation hole and its corresponding 
exchange-correlation potential of the 
electron gas are studied using the
random-phase approximation (RPA) \cite{pines1952,mahan,fetter-walecka}
to understand the salient features of the exchange-correlation hole and potential.
The long-term goal is to use the electron gas
results as a basis for a local density approximation in the spirit of density
functional theory \cite{hohenberg1964,kohn1965,jones1989,becke2014,jones2015}.

The theory section commences with a short summary of the exchange-correlation potential framework, 
which is outlined in detail in previous publications.
Formulas for the exchange-correlation hole 
and potential are then derived for the homogeneous electron gas.
An analysis of the sum rule and its consequences is presented in Sec. \ref{sec:Sumrule}
followed by
computational results in Sec. \ref{sec:Results} and a summary at the end. 

\section{Theory}

In the exchange-correlation potential formalism the equation of motion of the equilibrium 
zero temperature time-ordered
Green function is given by \cite{aryasetiawan2022a}
\begin{equation}
   \left(  i\frac{\partial}{\partial t}-h(r)-V_\mathrm{xc}(r,r';t)\right)
G(r,r^{\prime};t)= \delta(r-r^{\prime})\delta(t),
\label{eq:EOM}
\end{equation}
where
\begin{equation}
    h(r) = -\frac{1}{2}\nabla^2 +V_\mathrm{ext}(r)+V_\mathrm{H}(r),
\end{equation}
in which $V_\mathrm{ext}$ and $V_\mathrm{H}$ are the external field and 
the Hartree potential, respectively.
The Green function is defined according to
\cite{fetter-walecka}
\begin{equation}
iG(r,r^{\prime};t)=\langle T[\hat{\psi}(rt)\hat{\psi
}^{\dag}(r^{\prime}0)]\rangle ,
\end{equation}
where $r=(\mathbf{r},\sigma)$ labels both space and spin variables, $\hat
{\psi}(rt)$ is the Heisenberg field operator, $T$ is the
time-ordering symbol, and $\left\langle .\right\rangle $ denotes expectation
value in the ground state. 

The exchange-correlation potential $V_\mathrm{xc}$ is the Coulomb potential of 
the exchange-correlation hole $\rho_\mathrm{xc}$:
\begin{equation}
    V_\mathrm{xc}(r,r';t)=\int dr'' v(r-r'') \rho_\mathrm{xc}(r,r',r'';t).
    \label{eq:Vxc}
\end{equation}
The presence of the instantaneous Coulomb interaction implies that $t''=t$.
The exchange-correlation hole fulfills an important sum rule
\begin{equation}
    \int d^3r'' \rho_\mathrm{xc}(r,r',r'';t) = -\delta_{\sigma\sigma''} \theta(-t)
    \label{eq:sum-rule}
\end{equation}
and the following exact condition
\begin{equation}
    \rho_\mathrm{xc}(r,r',r''=r;t) =-\rho(r)
    \label{eq:exactcond}
\end{equation}
for \emph{any} $r$, $r'$ and $t$.

\subsection{General formula for the exchange-correlation hole}

From the definition of the exchange-correlation hole \cite{aryasetiawan2022a},
\begin{align}
    G^{(2)}&=\langle T[\hat{\rho}(3)\hat{\psi}(1)\hat{\psi}^\dag(2)]
    \nonumber\\
    &=i[\rho(3)+\rho_\mathrm{xc}(1,2,3)]G(1,2) ,
\end{align}
where $1=(r_1,t_1)$ etc. with $t_1=t_3$
and the relation \cite{hedin1965,aryasetiawan1998}
\begin{equation}
    G^{(2)}=i\rho(3)G(1,2)-\frac{\delta G(1,2)}{\delta\varphi(3)}, 
\end{equation}
an explicit formula for the exchange-correlation hole is given by \cite{aryasetiawan2022b}
\begin{align}
    \rho_\mathrm{xc}(1,2,3)=i\frac{\delta }{\delta\varphi(3)} \ln{G(1,2)},
\end{align}
where $\varphi$ is a probing field, 
which is set to zero after the functional derivative is taken. 
The exchange-correlation hole can thus be regarded as the linear response of $i\ln{G}$ with respect to
an external field.

From the identity 
\begin{equation}
    \delta G = -G (\delta G^{-1}) G
\end{equation}
and the equation of motion of the Green function, one obtains
\begin{align}
    \rho_\mathrm{xc}&(1,2,3)G(1,2)
    \nonumber\\
    &=i\int d4\,G(1,4)\left\{ \delta(3-4) +\frac{\delta V_\mathrm{H}(4)}{\delta\varphi(3)}
    \right\} G(4,2)
    \nonumber\\
    &+i\int d4d5\,G(1,4)\frac{\delta \Sigma(4,5)}{\delta\varphi(3)} G(5,2),
\end{align}
where $\Sigma$ is the self-energy. 
The first term on the right-hand side, $iG(1,3)G(3,2)$, will be referred to as the exchange contribution,
the second term involving $\frac{\delta V_\mathrm{H}}{\delta\varphi}$ as the 
linear response contribution, 
and the last term with $\frac{\delta \Sigma}{\delta\varphi}$ as the vertex correction.
The second and third terms together constitute the correlation contribution.
Within the RPA, the vertex correction is neglected.

The quantity in the curly brackets is the inverse dielectric 
function:
\begin{equation}
    \epsilon^{-1}(4,3)=\delta(4-3) +\frac{\delta V_\mathrm{H}(4)}{\delta\varphi(3)}.
\end{equation}
It is convenient for later purposes to define
\begin{align}
    K(4,3) &= \frac{\delta V_\mathrm{H}(4)}{\delta\varphi(3)} 
=\int d5\, v(4-5) \chi(5,3),
\label{def:K}
\end{align}
where $v$ is the Coulomb interaction and $\chi$ is the linear density response function
\begin{equation}
    \chi(5,3) = \frac{\delta\rho(5)}{\delta\varphi(3)}.
    \label{def:chi}
\end{equation}

Replacing $r_1\rightarrow r$, $r_2\rightarrow r'$, and $r_3\rightarrow r''$
and taking into account the fact that $t_1=t_3=t$ and $t_2=0$, one finds
\begin{align}
    &\rho_\mathrm{xc}(r,r',r'';t)G(r,r';t)
    =iG(r,r'';0^-)G(r'',r';t) 
    \nonumber\\
    &+ i\int dr_4dt_4\,G(r,r_4;t-t_4) K(r_4,r'';t_4-t)G(r_4,r';t_4).
    \label{eq:xc-hole}
\end{align}
The first term on the right-hand side yields the exchange hole whereas the second term
yields the correlation hole.
It is immediately clear that the exact condition in Eq. (\ref{eq:exactcond}) is
already fulfilled by the exchange hole implying that
\begin{equation}
    \rho_\mathrm{c}(r,r',r;t)=0.
    \label{eq:rhoc=0}
\end{equation}

\subsection{Interacting homogeneous electron gas}

The Green function of the paramagnetic
non-interacting homogeneous electron gas is given by
\begin{align}
    iG_0(r,r';t) &= \frac{1}{\Omega} \sum_{k> k_\mathrm{F}} 
    e^{i\mathbf{k}\cdot(\mathbf{r}-\mathbf{r}')} e^{-i\varepsilon_k t}\theta(t)
    \nonumber\\
    &-\frac{1}{\Omega} \sum_{k\leq k_\mathrm{F}} 
    e^{i\mathbf{k}\cdot(\mathbf{r}-\mathbf{r}')} e^{-i\varepsilon_k t}\theta(-t),
    \label{eq:G0}
\end{align}
where $\varepsilon_k=\frac{1}{2}k^2$, $k_\mathrm{F}$ is the Fermi wave vector,
and $\Omega$ is
the space volume. It is understood that $\sigma=\sigma'$.
For the homogeneous electron gas, it is convenient to introduce the variable 
$\mathbf{R}=\mathbf{r}'-\mathbf{r}$. In spherical coordinates the equation of motion becomes
\begin{equation}
   \left(  i\frac{\partial}{\partial t}-h(R)-V_\mathrm{xc}(R;t)\right)
\widetilde{G}(R,t)= \frac{1}{4\pi R}\delta(R)\delta(t),
\label{eq:EOMelgas}
\end{equation}
where
\begin{equation}
    h(R)=-\frac{1}{2}\frac{\partial^2}{\partial R^2},\qquad \widetilde{G}(R,t)=R \,G(R,t).
\end{equation}

Defining
\begin{equation}
    T(R,t)=\frac{1}{\widetilde{G}(R,t)}h(R)\widetilde{G}(R,t),
\end{equation}
the formal solution is given by
\begin{equation}
 {G} (R,t)= {G}(R,0)e^{-i\int_0^t dt' [T(R,t')+V_\mathrm{xc}(R,t')]} , 
\end{equation}
in which it is understood that ${G}(R,0)={G}(R,0^+)$ for $t>0$ 
and ${G}(R,0)={G}(R,0^-)$ for $t<0$.
In general, from the equation of motion in Eq. (\ref{eq:EOMelgas}) 
\begin{equation}
    i[G(R,0^+)-G(R,0^-)]=\frac{\delta(R)}{4\pi R^2}.
\end{equation}

\begin{figure}[tb]
\begin{center} 
\includegraphics[scale=0.5, viewport=7.5cm 6cm 25cm 15cm, clip,width=\columnwidth]
{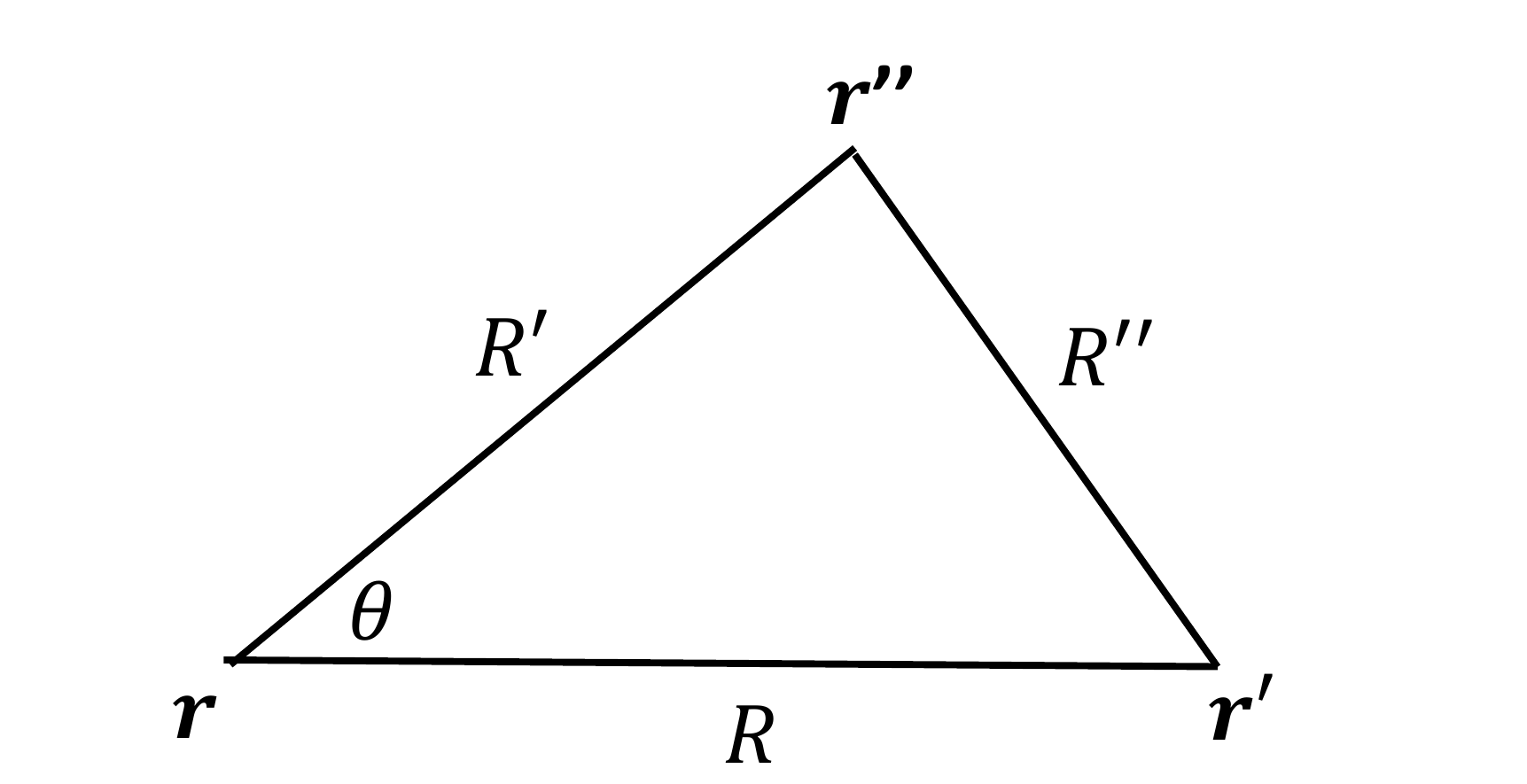}
\caption{Definition of the radial variables $R$, $R'$, and $R''$. They
are related to the angle $\theta$ by ${R''}^2=R^2-2RR'\cos{\theta} +{R'}^2$}.
\label{fig:Coordinates}%
\end{center}
\end{figure}

For the homogeneous electron gas, $\mathbf{r}$ may be chosen to be 
the origin, i.e., $\mathbf{r}=0$,
and one defines the variables $\mathbf{R}=\mathbf{r}'-\mathbf{r}$, 
$\mathbf{R}'=\mathbf{r}''-\mathbf{r}$, and 
$\mathbf{R}''=\mathbf{r}''-\mathbf{r}'$ as illustrated in
Fig. \ref{fig:Coordinates}.

\subsection{Exchange hole}
\label{sec:ExchangeHole}

From Eq. (\ref{eq:xc-hole}) the exchange hole is given by
\begin{align}
    \rho_\mathrm{x}(r,r',r'';t)G(r,r';t)
    &=iG(r,r'';0^-)G(r'',r';t).
    \label{eq:x-hole}
\end{align}
Unlike the static exchange hole \cite{slater1951} in quantum chemistry and
density functional theory, the exchange hole
in the present formalism is time dependent.

Using a non-interacting Green function as given 
in Eq. (\ref{eq:G0}) and considering the case $t<0$ one finds
\begin{align}
   & \rho_\mathrm{x}(r,r',r'';t<0)
    \sum_{k\leq k_F} 
    e^{i\mathbf{k}\cdot(\mathbf{r}-\mathbf{r}')} e^{-i\varepsilon_k t}
    \nonumber\\
    &=-\frac{1}{\Omega} \sum_{k'\leq k_F} 
    e^{i\mathbf{k}'\cdot(\mathbf{r}-\mathbf{r}'')}
    \times\sum_{k\leq k_F} 
    e^{i\mathbf{k}\cdot(\mathbf{r}''-\mathbf{r}')} e^{-i\varepsilon_k t}.
    \label{eq:rhox}
\end{align}
For $t>0$
\begin{align}
   & \rho_\mathrm{x}(r,r',r'';t>0)
    \sum_{k> k_F} 
    e^{i\mathbf{k}\cdot(\mathbf{r}-\mathbf{r}')} e^{-i\varepsilon_k t}
    \nonumber\\
    &=-\frac{1}{\Omega} \sum_{k'\leq k_F} 
    e^{i\mathbf{k}'\cdot(\mathbf{r}-\mathbf{r}'')}
    \times\sum_{k> k_F} 
    e^{i\mathbf{k}\cdot(\mathbf{r}''-\mathbf{r}')} e^{-i\varepsilon_k t}.
    \label{eq:rhox+}
\end{align}

Expressed in radial variables and the angle $\theta$ as explained in Fig. \ref{fig:Coordinates},
\begin{align}
    \rho_\mathrm{x}(R,R',\theta;t)
    &=iG_0(R',0^-)\frac{G_0(R'',t)}{G_0(R,t)} 
\end{align}
where $R''$ depends on $\theta$.
%

$G_0(R,t)$ is obtained from Eq. (\ref{eq:G0})
by performing the $\mathbf{k}$-integral over 
the solid angle, yielding
\begin{align}
    iG_0(R,t<0)
    &=-\frac{1}{2\pi^2}\frac{1}{R}\int_0^{k_\mathrm{F}} dk\, k\sin{(kR)}e^{-ik^2 t/2}, 
\end{align}
\begin{align}
    iG_0(R,t>0)
    &=\frac{1}{2\pi^2}\frac{1}{R}\int_{k_\mathrm{F}}^\infty dk\, k\sin{(kR)}
    e^{-ik^2 t/2} ,   
\end{align}
\begin{align}
    iG_0(R,0^-)
    &=-\frac{1}{2\pi^2}\frac{1}{R^3}\left[ \sin{(k_\mathrm{F}R)}
    -k_\mathrm{F}R\cos{(k_\mathrm{F}R)} \right] .
\end{align}
$G_0(R,t<0)$ can be expressed in terms of the complex error function or calculated
numerically using a standard quadrature. The calculation of $G_0(R,t>0)$, however,
needs more care and it is detailed in Appendix \ref{app:G0}.



\subsection{Correlation hole}

The linear response contribution to $\rho_\mathrm{xc}$ is given by the second term in Eq. (\ref{eq:xc-hole}).
%
%
Keeping in mind that $t_3=t_1=t$, $t_2=0$, one obtains
\begin{align}
    &i\int d4\,G(1,4) K(4,3)G(4,2)
    \nonumber\\
    &=i\int dr_4 dt_4 G(r-r_4,t-t_4) K(r_4-r'',t_4-t) 
    \nonumber\\
    &\times G(r_4-r',t_4).
\end{align}
The details of the calculations using $G=G_0$ are shown in Appendix \ref{app:rhoc} and the results
are given by
\begin{equation}
    \rho_\mathrm{c}(R,R',\theta;t<0)= \frac{A_1+A_2}{G_0(R,t<0)},
\end{equation}
\begin{equation}
    \rho_\mathrm{c}(R,R',\theta;t>0)= \frac{B_1+B_2}{G_0(R,t>0)}.
\end{equation}
$A_1$, $A_2$, $B_1$, and $B_2$ are functions of $R'$ and $R''$, 
and given by
\begin{align}
    A_1 &= \gamma(R',R'',t,0,t), \\
    A_2 &= \gamma(R'',R',t,0,0), \\
    B_1 &= \gamma(R',R'',0,t,0), \\
    B_2 &= \gamma(R'',R',0,t,-t),
\end{align}
where
\begin{align}
    &\gamma(R,R',t,t',t'')=
    \frac{1}{\Omega^2}\sum_{k\leq k_F}    
    e^{-i\mathbf{k}\cdot\mathbf{R}} e^{-i\varepsilon_{k}t}
    \nonumber\\
    &\times\sum_{k'> k_F}
    e^{i\mathbf{k}'\cdot\mathbf{R}'}e^{-i\varepsilon_{k'}t'}
M(|\mathbf{k}'-\mathbf{k}|,\varepsilon_{k'}-\varepsilon_{k},t'').
\end{align}


The quantity $M$ is given by
\begin{align}
    M(q,\omega,t) = 
    \int_0^\infty d\omega'\, L(q,\omega')
    \frac{-ie^{i\omega't}}
    {\omega'+\omega},
    \label{eq:Mqwt}
\end{align}
where
$L(q,\omega)$ is the spectral function of $K(q,\omega)$ defined in Eq. (\ref{def:K}):
\begin{align}
    K(k,\omega)=\int_{-\infty}^0 d\omega'\,\frac{L(k,\omega')}{\omega-\omega'-i\delta}
    +\int_0^{\infty}d\omega'\,\frac{L(k,\omega')}{\omega-\omega'+i\delta}.
\end{align}
$K(q,\omega)$ is symmetric in frequency but
$L(q,\omega)$ is anti-symmetric and related to $K$ as follows:
\begin{equation}
    L(k,\omega)=-\frac{1}{\pi}\mathrm{sign}(\omega) \mathrm{Im} K(k,\omega).
    \label{eq:Lkw}
\end{equation}

The correlation hole involves coupled
integrals over momenta below and above $k_\mathrm{F}$, which are difficult to calculate
analytically. They are six-dimensional integrals which cannot be easily performed with
standard quadratures. To make the computation feasible, a plasmon-pole approximation
for $L(k,\omega)$ is employed and 
described in the next section.

\subsubsection{Plasmon-pole approximation}
\label{sec:Plasmon-pole}

The plasmon dispersion of the homogeneous electron gas is given by \cite{fetter-walecka}
\begin{equation}
\Omega_{q}=\omega_{\text{p}}\left(  1+\frac{3}{10}\frac{k_{\text{F}}^{2}q^{2}%
}{\omega_{\text{p}}^{2}}+\text{ }...\right),
\end{equation}
where the plasmon frequency $\omega_\mathrm{p}$ in the long-wavelength limit is given by
\begin{equation}
    \omega_\mathrm{p}^2=4\pi \rho_0,
    \label{eq:plasmon}
\end{equation}
and $\rho_0$ is the electron gas density. 
The critical momentum, $q_{\text{c}}$, at which the plasmon starts to merge
into the electron-hole excitations is given by the crossing of the plasmon
dispersion with the line $\varepsilon_{q}+k_{\text{F}}q$
yielding%
\begin{equation}
q_{\text{c}}=\frac{1}{2a}\left(  \sqrt{1+4ac}-1\right)  k_{\text{F}},
\end{equation}
where%
\begin{equation}
a=\frac{1}{2}-\frac{3}{10c},\text{ \ \ }c=\frac{\omega_{\text{p}}}%
{k_{\text{F}}^{2}}.
\end{equation}

Great simplification results if a plasmon-pole approximation 
independent of $k$ is used for $L(k,\omega)$ defined in Eq. (\ref{eq:Lkw}):
\begin{equation}
    L(k,\omega)= \frac{\omega_\mathrm{p}}{2} \left[ 
    \delta(\omega-\omega_\mathrm{p}) - \delta(\omega+\omega_\mathrm{p}) \right], 
\end{equation}
which corresponds to
\begin{equation}
    K(q\rightarrow 0,\omega)=\frac{\omega_\mathrm{p}^2}{\omega^2-\omega_\mathrm{p}^2}.
\end{equation}
The approximation is valid for $k\leq q_\mathrm{c}$ and  
for $r_{s}=3$, $4$ and $5$, which cover most of the average valence
densities in real materials,
the critical momenta are
$q_{\text{c}}=0.86$, $0.82$, and $0.73$ $k_{\text{F}}$, respectively.

Within the plasmon-pole approximation the quantity $M$
defined in Eq. (\ref{eq:Mqwt}) becomes independent of momenta:
\begin{align}
    M(q,\omega,t) = \frac{\omega_\mathrm{p}}{2}
    \frac{-ie^{i\omega_\mathrm{p}t}}
    {\omega_\mathrm{p}+\omega}.
    \label{eq:Mpp}
\end{align}
Then
the coupling between $\mathbf{k}$ and $\mathbf{k}'$ is partially released.
Using
\begin{equation}
    \frac{1}{\Omega}\sum_{k}    
    e^{-i\mathbf{k}\cdot\mathbf{R}}= \frac{1}{2\pi^2 R}
    \int dk\, k \sin{(kR)} 
    \label{eq:sumk}
\end{equation}
yields within the plasmon-pole approximation
\begin{align}
    &\gamma^{PP}(R,R',t,t',t'')
    =\frac{-2i\omega_\mathrm{p}}{(2\pi)^4RR'}
    \int_0^{k_\mathrm{F}} dk\,k \sin{(kR)} e^{-i\varepsilon_k t}
    \nonumber\\
    &\qquad\qquad\times\int_{k_\mathrm{F}}^\infty dk'\,{k'} \sin{(k'R')}
     \frac{e^{-i\varepsilon_{k'} t'} e^{i\omega_\mathrm{p}t''}}
    {\omega_\mathrm{p}+\varepsilon_{k'}-\varepsilon_k}.
\end{align}

Since the plasmon-pole approximation
decouples the angular inter-dependence of $\mathbf{k}$ and $\mathbf{k}'$, it is expected to
impart error to the correlation contribution. To minimize error, the upper limit of the 
integration over $k'$ corresponding to unoccupied states is chosen so as to approximately
reproduce the static correlation hole \cite{wang1991}. 
This choice yields a value of $\approx 1.5 k_\mathrm{F}$
so that integration over unoccupied states arising from the correlation
contribution is restricted to between $k_\mathrm{F}$ and $1.5 k_\mathrm{F}$.

\subsection{Exchange-correlation potential}

By making a change of variable $\mathbf{R}'=\mathbf{r}''-\mathbf{r}$
the exchange-correlation potential in Eq. (\ref{eq:Vxc}) reduces to the first radial moment of the
spherical average of $\rho_\mathrm{xc}$:
%
%
%
\begin{equation}
V_\mathrm{xc}(r,r^{\prime};t)=\int dR'R'\text{ }\overline{\rho}%
_\mathrm{xc}(r,r^{\prime},R';t),
\label{Vxc1mom}
\end{equation}
where $\overline{\rho}_\mathrm{xc}(r,r^{\prime},R';t)$ for given
$r$, $r'$, and $t$ depends only on the radial
distance $R'=|\mathbf{r}''-\mathbf{r}|$,
\begin{equation}
\overline{\rho}_\mathrm{xc}(r,r^{\prime},R';t)
=\int d\Omega_{R'}\rho_\mathrm{xc}(r,r^{\prime
},\mathbf{r+R'};t).
\end{equation}

As can be seen from, e.g., Eqs. (\ref{eq:rhox}) and (\ref{eq:A1a}),
the spherical average of $\rho_{\text{xc}}$ for the electron gas
amounts to performing a solid-angle
integration
\begin{align}
\int d\Omega^{\prime\prime}e^{i(\mathbf{k}-\mathbf{k}^{\prime})\cdot
\mathbf{r}^{\prime\prime}} 
& =4\pi\frac{\sin(\Delta k\text{ }R^{\prime})}{\Delta k\text{ }R^{\prime}},
\end{align}
where $\Delta k=|\mathbf{k}-\mathbf{k}^{\prime}|$ and 
$R'=|\mathbf{r}^{\prime\prime}-\mathbf{r}|$, $\mathbf{r}=0$. 

\subsubsection{Exchange potential}

For $t<0$ the exchange hole is given by
\begin{align}
&\bar{\rho}_{\text{x}}(R,R^{\prime},t<0)iG_0(R,t)
\nonumber\\
&=\frac{1}{\Omega^{2}}%
\sum_{k,k^{\prime}\leq k_{\text{F}}}e^{-i\mathbf{k}\cdot\mathbf{r}^{\prime}%
}e^{-i\varepsilon_{k}t}\times4\pi\frac{\sin(\Delta k\text{
}R^{\prime})}{\Delta k\text{ }R^{\prime}}.
\end{align}

The exchange potential is the first moment of 
$\bar{\rho}_{\text{x}}$ in $R^{\prime}$:
\begin{align}
V_{\text{x}}(R,t<0)&=\frac{1}{iG_0(R,t)}
\frac{4\pi }{\Omega^{2}}\sum_{k,k^{\prime}\leq k_{\text{F}%
}}e^{-i\mathbf{k}\cdot\mathbf{R}}e^{-i\varepsilon_{k}t}
\nonumber\\
&\times \int
dR^{\prime}\frac{\sin(\Delta k\text{ }R^{\prime})}{\Delta k}.%
\end{align}
Consider the integral over $R'$ with positive $\alpha\rightarrow 0$:
\begin{equation}
\lim_{\alpha\rightarrow 0}
\int_{0}^{\infty}dR^{\prime}\sin(\Delta k\text{ }R^{\prime})
e^{-\alpha R'}
 =\frac{1}{\Delta k}.
\end{equation}
One then finds
\begin{equation}
V_{\text{x}}(R,t<0)=\frac{1}{iG_0(R,t)}\frac{4\pi}{\Omega^{2}}\sum_{k,k^{\prime}\leq
k_{\text{F}}}e^{-i\mathbf{k}\cdot\mathbf{R}}e^{-i\varepsilon_{k}t}\frac
{1}{(\Delta k)^{2}}.
\end{equation}
The integral over $k^{\prime}$ is given by%
\begin{align}
f(k)  & =\frac{1}{\Omega}\sum_{k^{\prime}\leq k_{\text{F}}}\frac{1}{(\Delta
k)^{2}}
\nonumber\\
& =\frac{1}{4\pi^{2}k}\int_{0}^{k_{\text{F}}}dk^{\prime}k^{\prime}%
\ln\left\vert \frac{k+k^{\prime}}{k-k^{\prime}}\right\vert ,
\label{eq:fk}
\end{align}
which can be performed analytically 
yielding
\begin{align}
    f(k)&=\frac{k_\mathrm{F}}{2\pi^2} F(k/k_\mathrm{F}),
\end{align}
where
\begin{equation}
    F(x)= 
    \frac{1}{2}+\frac{1-x^{2}}{4x} \ln\left\vert \frac{1+x}{1-x}\right\vert .
\end{equation}
This function is the same as the one appearing in the static Hartree-Fock theory for the
electron gas \cite{ashcroft}.
More explicitly as a function of $k$,
\begin{align}
    f(k)
    &=\frac{k_\mathrm{F}}{2\pi^2}
    \left(
    \frac{1}{2}+\frac{k^2_\mathrm{F}-k^2}{4k_\mathrm{F}k} \ln\left\vert
    \frac{k_\mathrm{F}+k}{k_\mathrm{F}-k}\right\vert 
    \right).
\end{align}
There remains the integral over $\mathbf{k}$ which reduces to a one-dimensional 
integral over the radial $k$:

\begin{align}
V_{\text{x}}(R,t<0)&=\frac{1}{iG_0(R,t)}
\nonumber\\ 
&\quad\times \frac{2}{\pi R}\int_{0}%
^{k_{\text{F}}}dk\text{ }k\sin(kR)\text{ }e^{-i\varepsilon_{k}t}f(k).
\label{eq:Vx-}
\end{align}
For $t>0$ the result is given by 
\begin{align}
V_{\text{x}}(R,t>0)&=-\frac{1}{iG_0(R,t)}
\nonumber\\ 
&\quad\times \frac{2}{\pi R}\int_{k_\text{F}}%
^{\infty}dk\text{ }k\sin(kR)\text{ }e^{-i\varepsilon_{k}t}f(k).
\label{eq:Vx+}
\end{align}

\subsubsection{Correlation potential}
A similar procedure as for the exchange potential can be applied to 
the correlation potential using $A_1$, $A_2$, $B_1$, and $B_2$, 
given in Eqs. (\ref{eq:A1M}-\ref{eq:B2M}) in Appendix \ref{app:rhoc}.

The result is given by
\begin{align}
V_{\text{c}}(R,t<0)&=\frac{C_1+C_2}{G_0(R,t<0)},
\end{align}
\begin{align}
V_{\text{c}}(R,t>0)&=\frac{D_1+D_2}{G_0(R,t>0)},
\end{align}
where
\begin{align}
    C_1&=\Gamma(0,R,t,0,t),\\
    C_2&=\Gamma(R,0,t,0,0),\\
    D_1&=\Gamma(0,R,0,t,0),\\
    D_2&=\Gamma(R,0,0,t,-t),
\end{align}
and
\begin{align}
&\Gamma(R,R',t,t',t'')=
\frac{4\pi }{\Omega^2}\sum_{k\leq k_\mathrm{F}}  
    e^{-i\varepsilon_{k}t}e^{-i\mathbf{k}\cdot\mathbf{R}}
\nonumber\\ 
&\times   \sum_{k'> k_\mathrm{F}}
    e^{-i\mathbf{k}'\cdot\mathbf{R}'}
    e^{-i\varepsilon_{k'}t'}
    \times\frac{M(|\mathbf{k}'-\mathbf{k}|,\varepsilon_{k'}-\varepsilon_{k},t'')}
    {|\mathbf{k}'-\mathbf{k}|^2}.
    \label{eq:Gamma}
\end{align}

According to Eq. (\ref{eq:Mpp}), within the plasmon-pole approximation,
\begin{equation}
  M(|\mathbf{k}'-\mathbf{k}|,\varepsilon_{k'}-\varepsilon_{k},t'')
  =\frac{\omega_\mathrm{p}}{2}
    \frac{-ie^{i\omega_\mathrm{p}t''}}{\omega_\mathrm{p}+\varepsilon_{k'}-\varepsilon_{k}},
\end{equation}
which partially decouples the interdependence of $\mathbf{k}$ and $\mathbf{k}'$,
allowing for analytical integration over the solid angles of both variables, yielding
\begin{align}
    C_1&=P_1(R,t,0,t),\\
    C_2&=P_2(R,t,0,0),\\
    D_1&=P_1(R,0,t,0),\\
    D_2&=P_2(R,0,t,-t),
\end{align}
where
\begin{align}
P_1(R,t,t',t'')
&=
\frac{-i\omega_\mathrm{p}e^{i\omega_\mathrm{p}t''}}{4\pi^3 R}\int_0^{k_\mathrm{F}}dk\
\int_{k_\mathrm{F}}^\infty dk'\,k\sin{(k'R)}
\nonumber\\
    &\times
         \frac{e^{-i\varepsilon_{k}t}e^{-i\varepsilon_{k'}t'}}
    {\omega_\mathrm{p}+\varepsilon_{k'}-\varepsilon_{k}}
    \ln{\left\vert \frac{k+k'}{k-k'}\right\vert},
\end{align}
\begin{align}
P_2(R,t,t',t'')
&=
\frac{-i\omega_\mathrm{p}e^{i\omega_\mathrm{p}t''}}{4\pi^3 R}\int_0^{k_\mathrm{F}}dk\
\int_{k_\mathrm{F}}^\infty dk'\,k'\sin{(kR)}
\nonumber\\
    &\times
    \frac{e^{-i\varepsilon_{k}t}e^{-i\varepsilon_{k'}t'}}
    {\omega_\mathrm{p}+\varepsilon_{k'}-\varepsilon_{k}}
         \ln{\left\vert \frac{k+k'}{k-k'}\right\vert}.
\end{align}
Due to the use of the plasmon-pole approximation, the upper limit of the integral over $k'$
is restricted to $1.5 k_\mathrm{F}$ to reproduce approximately the static
correlation hole, as described earlier in Sec. \ref{sec:Plasmon-pole}.

\subsection{Sum rule}
\label{sec:Sumrule}

In this section, the sum rule and its consequences are discussed and
a simple vertex approximation respecting the sum rule is proposed.
The results and conclusions reached 
in this section are quite general and supported by the electron gas results.

\subsubsection{Exchange hole}

One has for a non-interacting $G$ and $t<0$
\begin{align}
    i\int dr'' G_0(r,r'';0^-)G_0(r'',r';t<0)
    &=-G_0(r,r';t<0),
\end{align}
which can be shown as follows:
\begin{align}
    &i\int dr'' G_0(r,r'';0^-)G_0(r'',r';t<0)
    \nonumber\\
    &=-i\int dr'' \sum_{k\leq k_\mathrm{F}} \varphi_k(r)\varphi_k^*(r'')
    \sum_{k'\leq k_\mathrm{F}} \varphi_{k'}(r'')\varphi_{k'}^*(r')
    e^{-i\varepsilon_{k'}t}
    \nonumber\\
    &=-i\sum_{k\leq k_\mathrm{F}} \varphi_k(r)\varphi_k^*(r')
    e^{-i\varepsilon_{k}t}
    \nonumber\\
    &=-G_0(r,r';t<0).
\end{align}
It is also quite clear that
\begin{equation}
    i\int dr'' G_0(r,r'';0^-)G_0(r'',r';t>0)=0.
\end{equation}
It then follows from Eq. (\ref{eq:x-hole}) that the exchange hole fulfills the sum
rule when $G_0$ is used:
\begin{equation}
    \int dr'' \rho_\mathrm{x}(r,r',r'';t<0)= -1.
\end{equation}
Explicitly for the electron gas, it follows from Eq. (\ref{eq:rhox}) 
that the sum rule 
for $t<0$ is fulfilled by the exchange hole since
\begin{equation}
    \int d^3r'' e^{i(\mathbf{k}-\mathbf{k}')\cdot \mathbf{r}''}
    =(2\pi)^3\delta(\mathbf{k}-\mathbf{k}').
\end{equation}
The sum rule for $t>0$ is zero since $\mathbf{k}\neq \mathbf{k}'$ as can be seen
from Eq. (\ref{eq:rhox+}).

In general
\begin{equation}
      -i\int dr'' G(r,r'';0^-)G(r'',r';t<0) \neq G(r,r';t<0)
\end{equation}
and also
\begin{equation}
    i\int dr'' G(r,r'';0^-)G(r'',r';t>0)\neq 0.
\end{equation}
unless $G=G_0$.
This implies that if only the exchange part is considered, neglecting the correlation
and vertex terms, then in general
the sum rule is not fulfilled when a renormalized $G$ is used.

\subsubsection{Correlation hole}

It is also evident that the correlation part of the exchange-correlation 
hole gives no contribution to
the sum rule. The reason for this can be seen by considering 
the change in the charge density under a perturbation:
\begin{equation}
   \delta\rho(1)=\int d2\, \chi(1,2) \delta\varphi(2),
   \label{eq:deltarho}
\end{equation}
where $\chi$ is the linear density response function as defined in Eq. (\ref{def:chi}).
A constant perturbation, $\delta\varphi=1$, does not alter the density so that
\begin{equation}
   \int d2\, \chi(1,2) = 0. 
   \label{eq:chi0}
\end{equation}
This property is fulfilled by the RPA response function calculated using $G_0$ as shown
below.

It is interesting to observe that in the case of the electron gas, 
it can be seen explicitly from Eq. (\ref{eq:A1a}) 
in Appendix \ref{app:rhoc}
that the integral of $A_1$ over $r''$ is zero since $k\neq k'$ and 
the same conclusion holds for $A_2$, $B_1$, and $B_2$.
Hence the sum-rule is fulfilled, irrespective
of the approximation used for $K(q,\omega)$. 

Since the response function can be expanded in powers of the polarization $P$,
\begin{equation}
    \chi = P+ PvP + ...
\end{equation}
it follows that if the polarization function fulfills
\begin{equation}
    \int d2 P(1,2) = 0
    \label{eq:P0}
\end{equation}
then Eq. (\ref{eq:chi0}) is satisfied.

The polarization in the RPA is given by
\begin{equation}
    P(r,r';t) = -iG(r,r';t) G(r',r;-t).
    \label{eq:PRPA}
\end{equation}
If a non-interacting $G$ is used, then Eq. (\ref{eq:P0}) and consequently 
Eq. (\ref{eq:chi0}) are fulfilled. This can be understood as follows.
It can be shown that
\begin{equation}
    \int dr'' G_0(r,r'';t)G_0(r'',r';-t) = 0
    \label{eq:G0G0}
\end{equation}
by using the definition of $G_0$:
\begin{align}
    &\int dr'' G_0(r,r'';t)G_0(r'',r';-t)
    \nonumber\\
    &=\theta(t)\int dr'' \langle \hat{\psi}(rt)\hat{\psi}^\dag(r'')\rangle 
                \langle \hat{\psi}^\dag(r')\hat{\psi}(r'',-t)\rangle 
    \nonumber\\
    &+\theta(-t)\int dr'' \langle \hat{\psi}^\dag(r'')\hat{\psi}(rt)\rangle 
                \langle \hat{\psi}(r'',-t)\hat{\psi}^\dag(r')\rangle .
\end{align}
Since the expectation value is taken with respect to a single Slater determinant,
the integral over $r''$ amounts to
\begin{equation}
    \int dr'' \varphi^*_k(r'')\varphi_{k'}(r'') = 0,
\end{equation}
where $\{\varphi_k\}$ is a set of orbitals defining the single Slater determinant,
and if $k$ corresponds to an occupied state then $k'$ corresponds to an unoccupied state
and vice versa.

However, if a renormalized $G$ is used to calculate $P$ in the RPA as in 
Eq. (\ref{eq:PRPA}), 
the conditions in Eq. (\ref{eq:P0}) and consequently 
Eq. (\ref{eq:chi0}) are no longer fulfilled in general.
Thus, with a renormalized $G$,
both the exchange and correlation terms would violate the sum rule, and
it seems unlikely that the sum of these two terms would fulfill the sum rule.
This might explain why it is not favorable to use a renormalized $G$ in the
RPA and consequently in the $GW$ approximation \cite{hedin1965,aryasetiawan1998}. 
It also implies that inclusion of the vertex
$\delta\Sigma/\delta\varphi$ would restore the sum rule. Hence to preserve the sum rule, 
the use of a renormalized $G$ should be accompanied by inclusion of the vertex.

\subsubsection{A simple vertex correction}

The preceding consideration suggests a scheme 
for an approximate vertex correction that would preserve the sum rule.
Consider the following screened-exchange self-energy,
\begin{equation}
    \Sigma(1,2)= iG_0(1,2)W_0(1,2),
\end{equation}
where $W_0(1,2)$ is an instantaneous interaction,
\begin{equation}
W_0(1,2)=W_0(r_1,r_2)\delta(t_1-t_2),    
\end{equation}
with $W_0(r_1,r_2)$ being some screened interaction, e.g., 
the Thomas-Fermi screened interaction or the static screened interaction
calculated within RPA.
By using $G_0$ throughout, the vertex correction is given by
\begin{align}
    &\frac{\delta\Sigma(1,2)}{\delta\varphi(3)}
    =i\frac{\delta G_0(1,2)}{\delta\varphi(3)} W_0(1,2)
    \nonumber\\
&=-i\int d4d5\,G_0(1,4)\frac{\delta G^{-1}_0(4,5)}{\delta\varphi(3)} G_0(5,2) W_0(1,2),
\end{align}
where the identity $\delta G=-G(\delta G^{-1}) G$ has been utilised and
$\frac{\delta W_0}{\delta\varphi}=0$ since it is assumed that $W_0$ is fixed.
In the presence of the probing field $\varphi$, $G_0$ fulfills the equation of
motion
\begin{equation}
   \left(  i\frac{\partial}{\partial t_1}-h(1) -\varphi(1)\right)
G_0(1,2)= \delta(1-2),
\label{eq:EOM0}
\end{equation}
so that
\begin{equation}
    \frac{\delta G^{-1}_0(4,5)}{\delta\varphi(3)}=-\delta(4-5)
    \left[\delta(3-4) + \frac{\delta V_\mathrm{H}(4)}{\delta\varphi(3)}
    \right].
\end{equation}
This leads to
\begin{align}
    \frac{\delta\Sigma(1,2)}{\delta\varphi(3)}
&=iG_0(1,3)G_0(3,2) W_0(1,2)
\nonumber\\
&+i\int d4\,G_0(1,4)\frac{\delta V_\mathrm{H}(4)}{\delta\varphi(3)} G_0(4,2) W_0(1,2),
\label{eq:vertex}
\end{align}
and hence according to Eqs. (\ref{eq:chi0}) and (\ref{eq:G0G0}),
\begin{equation}
    \int d3 \frac{\delta\Sigma(1,2)}{\delta\varphi(3)}=0,
\end{equation}
noting that $t_1=t_2$ due to the instantaneous interaction.
It follows that the vertex in Eq. (\ref{eq:vertex}) preserves the sum rule.

\subsection{A model Green function for the interacting electron gas}
\label{sec:ModelG}

To assist in analyzing the results of the calculations of $V_\mathrm{xc}$ of the electron gas,
a model Green function has been constructed.
A physically motivated model for the Green function of the interacting electron gas
is given by the following:
\begin{align}
    G(R,t<0) &=
    \frac{i}{\Omega}\sum_{k\leq k_\mathrm{F}}
    \left[Z_k+(1-Z_k)e^{i\omega_k t }\right]
    \nonumber\\
    &\qquad\qquad \times e^{-iE_k t } e^{i\mathbf{k}\cdot\mathbf{R}},
    \label{eq:ModelGh}
\end{align}
\begin{align}
    G(R,t>0) &=
   - \frac{i}{\Omega}\sum_{k> k_\mathrm{F}}
    \left[Z_k+(1-Z_k)e^{-i\omega_k t }\right]
    \nonumber\\
    &\qquad\qquad\times e^{-iE_k t } e^{i\mathbf{k}\cdot\mathbf{R}},
    \label{eq:ModelGe}
\end{align}
where $E_k$ is the quasiparticle energy, $Z_k$ is the quasiparticle
renormalization factor, 
and $\omega_k$ is the plasmon energy. 
To allow for analytic derivation of the
exchange-correlation potential, $Z_k$ and $\omega_k$ are 
assumed to be independent of $k$
and $E_k$ is taken to be a renormalized
free-electron gas dispersion:
\begin{equation}
    Z_k=Z,\qquad \omega_k=\omega_\mathrm{p},\qquad E_k=\alpha\varepsilon_k
    =\frac{\alpha}{2}k^2.
    \label{eq:Zk}
\end{equation}
For an electron gas of density $\rho_0$ the plasmon energy is given in Eq. (\ref{eq:plasmon}).

The model Green function can be improved by including the possibility of having
spectral weight above or below the Fermi level for $k\leq k_\text{F}$ 
or $k > k_\text{F}$, respectively:
\begin{align}
    &G(R,t<0) = \frac{i}{\Omega}\sum_{k\leq k_\mathrm{F}} (C_1 +C_2 +C_3)
    e^{i\mathbf{k}\cdot\mathbf{R}},
\end{align}
\begin{align}
    &G(R,t>0) =-\frac{i}{\Omega}\sum_{k> k_\mathrm{F}} (D_1 +D_2 +D_3)
    e^{i\mathbf{k}\cdot\mathbf{R}},
\end{align}
where
\begin{align}
    C_1 &= Z e^{-iE_k t },\\
    C_2 &= (1-\beta^-_k)(1-Z)e^{-i(E_k-\omega_\mathrm{p}) t },\\
    C_3 &= \beta^-_k(1-Z)e^{-i(E_\mathrm{F}+\omega_\mathrm{p}) t },
\end{align}
\begin{align}
    D_1 &= Z_\mathrm{p} e^{-iE_k t },\\
    D_2 &= (1-\beta^+_k)(1-Z_\mathrm{p})e^{-i(E_k+\omega_\mathrm{p}) t },\\
    D_3 &= \beta^+_k(1-Z_\mathrm{p})e^{-i(E_\mathrm{F}-\omega_\mathrm{p}) t },
\end{align}
in which
\begin{align}
 \beta^-_k&=\frac{k}{2k_\mathrm{F}}\theta(k_\mathrm{F}-k) ,\\
 \beta^+_k&=\frac{1}{2}\frac{k_\mathrm{max}-k}{k_\mathrm{max}-k_\mathrm{F}}
 \theta(k_\mathrm{max}-k) .
\end{align}
For $k>k_\mathrm{max}$, the dispersion is taken to be that of the free-electron gas for otherwise
the integration over $k$ to infinity would not converge. It means physically that high-energy
electrons are free since they do not experience exchange and correlations.
To conserve electronic charge, the weight above the Fermi level coming from states smaller than 
$k_\mathrm{F}$ is equated to the weight below the Fermi level coming from states larger than 
$k_\mathrm{F}$, yielding an upper limit, neglecting the $k$-dependence of $\beta^\pm$,
\begin{align}
k_\mathrm{max}\approx \left( 1+\frac{1-Z}{1-Z_\mathrm{p}}\right)^{1/3}k_\mathrm{F}    
\end{align}
above which the spectrum
does not have weight below the Fermi level.

For $t\neq 0$,
the exchange-correlation potential can be obtained from the equation of motion:
\begin{align}
    V_\mathrm{xc}(R,t)& =\frac{1}{G(R,t)}\left[i\partial_t -h\right] G(R,t).
    \label{eq:VxcR}
\end{align}
Since
\begin{equation}
 h \exp{(i\mathbf{k}\cdot\mathbf{R})}=\frac{k^2}{2} \exp{(i\mathbf{k}\cdot\mathbf{R})}
\end{equation}
one finds for $t<0$
\begin{align}
  & \left[i\partial_t -h\right] G(R,t)
  \nonumber\\
  &= i\frac{1}{\Omega}\sum_{k\leq k_\mathrm{F}} (A_1+A_2+A_3)
  e^{i\mathbf{k}\cdot\mathbf{R}},
  \label{eq:derGh}
\end{align}
where
\begin{align}
    A_1 &= Z(E_k-\varepsilon_k)e^{-iE_k t }, \\
    A_2 &= (1-\beta^-_k)(1-Z)(E_k-\varepsilon_k-\omega_\mathrm{p})
    e^{-i(E_k-\omega_\mathrm{p}) t },\\
    A_3 &= \beta^-_k(1-Z)(E_\mathrm{F}-\varepsilon_k+\omega_\mathrm{p})
    e^{-i(E_\mathrm{F}+\omega_\mathrm{p}) t }.
\end{align}
For $t>0$
\begin{align}
  & \left[i\partial_t -h\right] G(R,t)
  \nonumber\\
  &=  -i \frac{1}{\Omega} \sum_{k>k_\mathrm{F}} (B_1+B_2+B_3)
    e^{i\mathbf{k}\cdot\mathbf{R}},
\end{align}
where
\begin{align}
    B_1 &= Z_\mathrm{p}(E_k-\varepsilon_k)e^{-iE_k t }, \\
    B_2 &= (1-\beta^+_k) (1-Z_\mathrm{p})(E_k-\varepsilon_k+\omega_\mathrm{p}) 
    e^{-i(E_k+\omega_\mathrm{p}) t },\\
    B_3 &= \beta^+_k (1-Z_\mathrm{p})(E_\mathrm{F}-\varepsilon_k-\omega_\mathrm{p})
    e^{-i(E_\mathrm{F}-\omega_\mathrm{p}) t }.
\end{align}

\section{Results}
\label{sec:Results}

The results shown in this section are all expressed in atomic units (a.u.) and
correspond to $r_s=4$. Two pairs of
representative times,
$t=\pm 34.75$ corresponding to $1/\bar{\varepsilon}$, where $\bar{\varepsilon}$
is the center of the occupied band, and $t=\pm 4.62$ corresponding to the inverse of
the plasmon energy,
$1/\omega_\mathrm{p}$, have been chosen for illustrations.
Additional times are also considered as appropriate.

\subsection{Angular dependence of the exchange-correlation hole}

\begin{figure}[t]
\begin{center} 
\includegraphics[scale=0.6, viewport=3cm 8cm 17cm 20cm, clip,width=\columnwidth]
{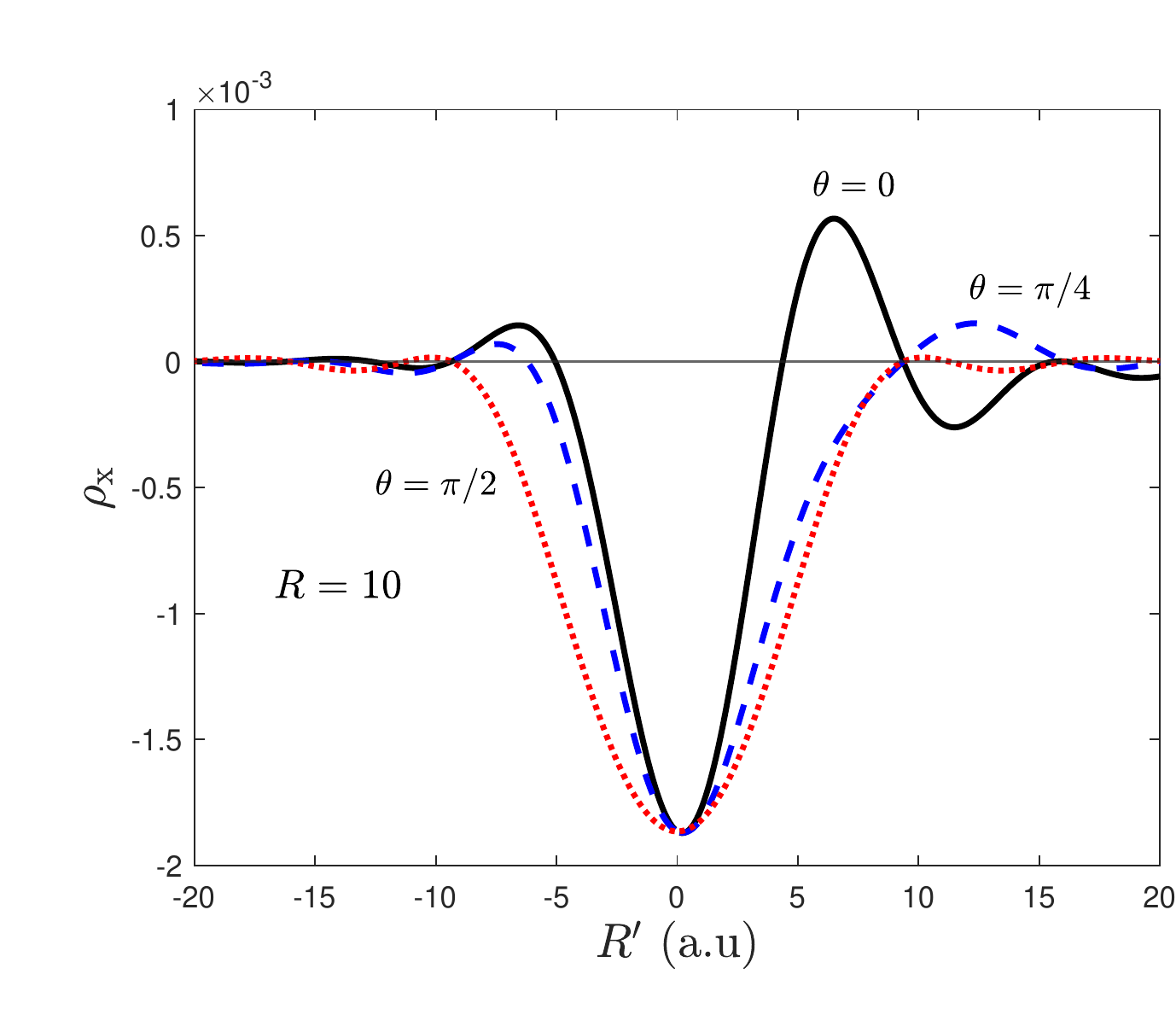}
\hfill
\includegraphics[scale=0.6, viewport=3cm 8cm 17cm 20cm, clip,width=\columnwidth]
{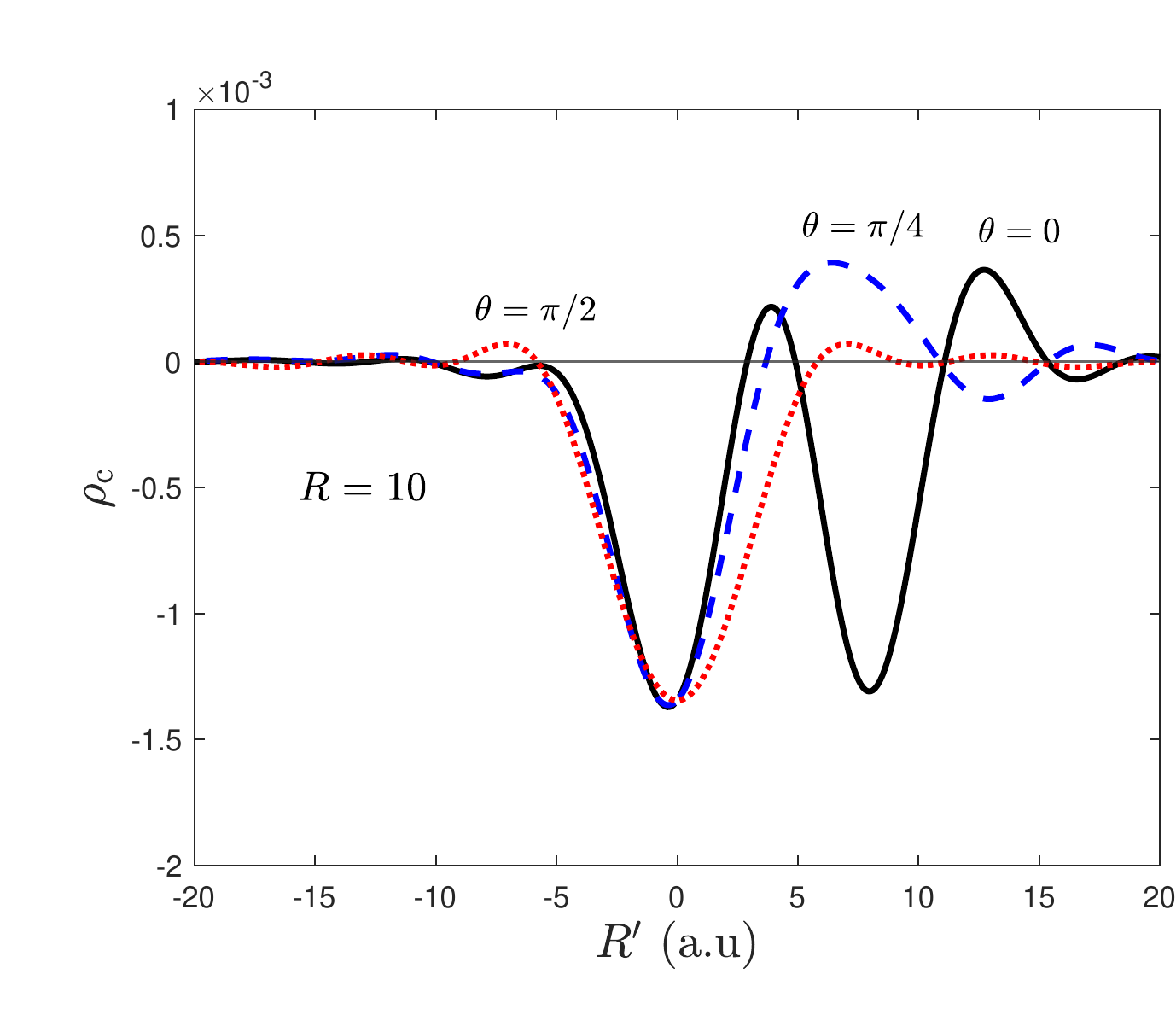}
\caption{The real part of the exchange hole (top) and the correlation hole (bottom)
for $\theta=0$ (solid black), $\pi/4$ (dashed blue), and $\pi/2$ (dotted red) for the case
$R=10$ and $t=-34.75$.
}
\label{fig:rhoxcR10Theta}%
\end{center}
\end{figure}

In Fig. \ref{fig:rhoxcR10Theta} the real part of the exchange holes for the case
$R=10$, $t=-34.75$, and for three
different angles $\theta=0,\,\pi/4,\,\pi/2$ as defined in Fig. \ref{fig:Coordinates}
are shown. These density fluctuations are due to exchange-only interaction
and correspond to the case
when a hole created at $R=0$ at $t=-34.75$ is to be annihilated later at $R=10$ at $t=0$.
It can be seen that there are more fluctuations for $\theta=0$ representing
the direction along the line connecting the location
where the hole is created at $R=0$
and the location where the hole is annihilated at $R=10$. The fluctuations are stronger
in the direction towards $R=10$ from the origin than in the opposite direction, which
reflect the preponderance of the hole presence and its effects in the former direction.
Indeed, the density fluctuations along the direction perpendicular to
the line joining $R=0$ and $R=10$ ($\theta=\pi/2$) 
are found to be the least whereas for the direction
$\theta=\pi/4$ they are somewhere in between those of $\theta=0$ and $\pi/2$.
In accordance with discussion in the paragraph following
Eq. (\ref{eq:xc-hole}), the exchange hole already fulfills the exact condition in
Eq. (\ref{eq:exactcond}), i.e., $\rho_\mathrm{x}(0)=-\rho_0$, where
$\rho_0$ the homogeneous electron gas density.

A similar behavior can be observed for the correlation holes shown in 
Fig. \ref{fig:rhoxcR10Theta}. These density fluctuations arise from the classical
Coulomb interaction in response to the creation of a hole and its accompanying
exchange hole. The fluctuations are 
particularly strong along the direction towards $R=10$ from the origin. According 
to Eq. (\ref{eq:rhoc=0}), the correlation hole at the origin should vanish, which
is clearly not the case here. There are two possible explanations for this discrepancy,
one being the use of the plasmon-pole approximation and another being the neglect
of the vertex term. One may speculate that the vertex term is needed
even when the exact response function is used, but to answer this issue in the definitive,
calculations employing the full RPA response function are required.
It is, however, to be noted that discrepancy may not be as severe as it might appear at
first sight because what enters into the exchange-correlation potential is the first
moment of the spherical average of the exchange-correlation hole so that contributions
at small $R$ are significantly cut down.
Numerical calculations of the sum rule show that in accordance with theory the
exchange hole alone fulfills the sum rule for $t<0$ and the correlation hole integrates
to zero whereas for $t>0$ both the exchange and the correlation holes integrate to zero.

\subsection{Time dependence of the exchange-correlation hole}

\begin{figure}[t]
\begin{center} 
\includegraphics[scale=0.6, viewport=3cm 8cm 17cm 20cm, clip,width=\columnwidth]
{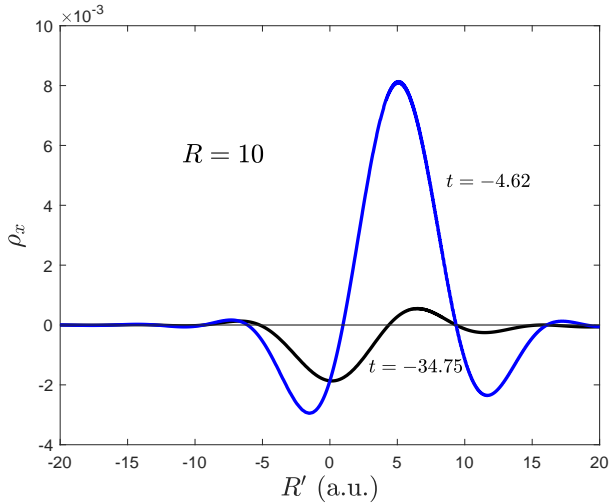}
\caption{
The real part of the exchange hole for $t=-4.62$ and $-34.75$ and the case $R=10$, $\theta=0$.
}
\label{fig:rhoxR10}%
\end{center}
\end{figure}

To illustrate the time dependence of the exchange-correlation hole, the exchange holes
for two representative times $t=-4.62$ and $t=-34.75$ are shown in Fig. \ref{fig:rhoxR10}
for the case $R=10$ and $\theta=0$. A similar result is found for the correlation hole.
For a short time scale in which the hole is annihilated at a location well 
separated from the location where it was created,
the fluctuations are much stronger than those of a long period.
As a guide, it is useful to consider the limit of $R=0$ and $t=0^-$, which yields
the static exchange-correlation hole. There is a correlation between the spatial
separation $R$ and the time period,
which determines the behavior of the exchange-correlation hole.
Within a short
period of time but with large $R$, 
the system may not have sufficient time to relax so that the density
fluctuations arising from a sudden creation of a hole at $t=-4.62$ have not stabilized.
These large density fluctuations for small $\theta$, however, contribute little
to the exchange-correlation potential since
only the spherical average of the exchange-correlation hole is needed and
when taking the spherical average each angular 
contribution is multiplied by $\sin{\theta}$.

\begin{figure}[t]
\begin{center} 
\includegraphics[scale=0.6, viewport=4cm 8cm 17cm 20cm, clip,width=\columnwidth]
{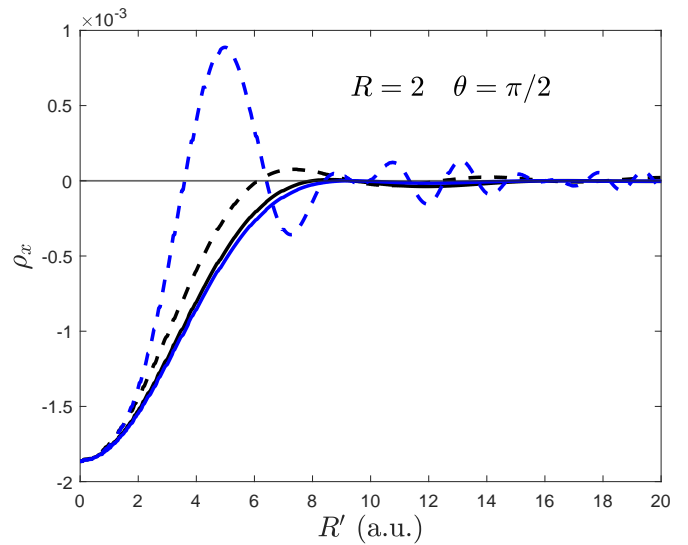}
\caption{
The real part of the exchange hole for $t=\pm 4.62$ (blue) and $\pm 34.75$ (black) 
for $R=2$, $\theta=\pi/2$.
The solid and dashed lines correspond to $t<0$ and $t>0$, respectively.
}
\label{fig:rhoxR2T4-34}%
\end{center}
\end{figure}

On the other hand, for a relatively short spatial separation $R=2$, the time-dependence
has a much less influence on the exchange hole, as can be seen in 
Fig. \ref{fig:rhoxR2T4-34}.
The difference between the exchange holes for $t=-4.62$
and $-34.75$ is much less pronounced than for the case of $R=10$.
The case of small $R$ and small $t$ approaches the static exchange-correlation hole limit
of $R=0$ and $t=0^-$.

For $t>0$ corresponding to an electron addition, the exchange hole exhibits a more
oscillatory behavior than for $t<0$. This can be understood from the exact sum rule
which dictates that the exchange hole must integrate to zero for $t>0$. For larger $t$,
however, the oscillations become damped as the system stabilizes itself.


\subsection{Spherical average of the exchange-correlation hole}

\begin{figure}[H]
\begin{center} 
\includegraphics[scale=0.6, viewport=3.2cm 8cm 17.8cm 20cm, clip,width=\columnwidth]
{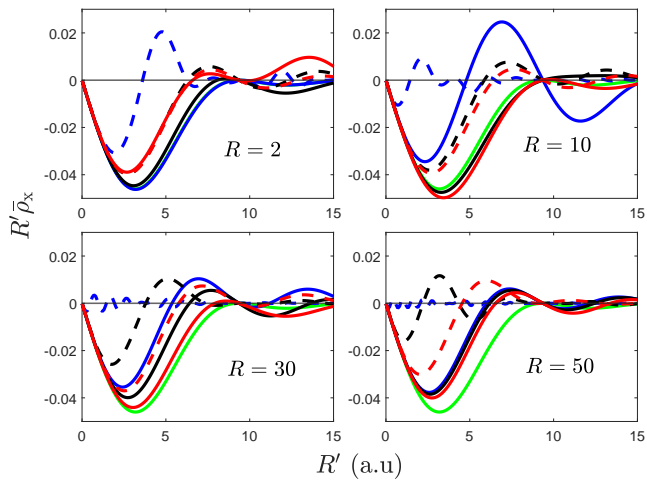}
\hfill
\includegraphics[scale=0.6, viewport=3.2cm 8cm 17.8cm 20cm, clip,width=\columnwidth]
{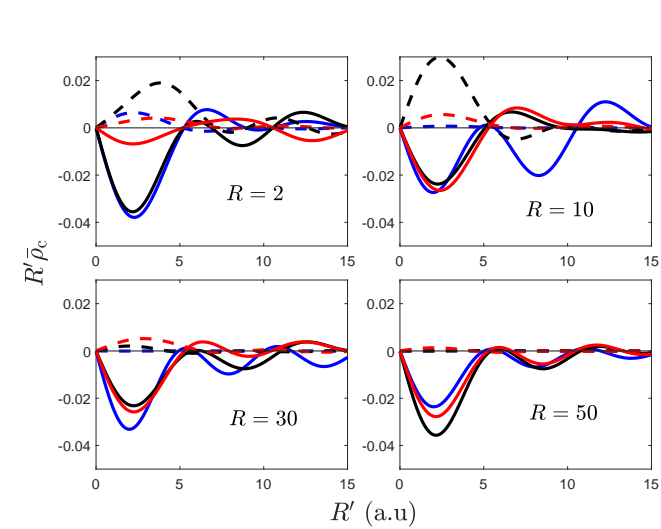}
\includegraphics[scale=0.6, viewport=3.2cm 8cm 17.8cm 20cm, clip,width=\columnwidth]
{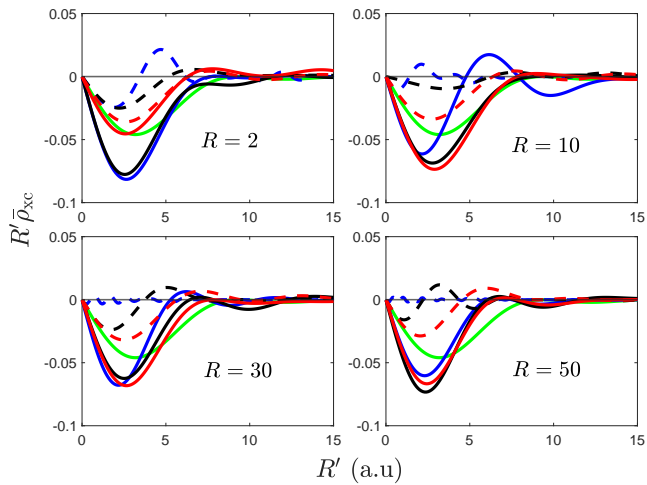}
\caption{
The real part of the spherical average of the 
exchange hole (upper panel), the correlation hole (middle panel), and
the exchange-correlation hole (lower panel)
multiplied by $R'$ for $t=\pm 4.62$ (blue), $\pm 34.75$ (black), $\pm 69.5$ (red),
and $R=2,10,30,50$.
The solid and dashed curves correspond to $t<0$ and $t>0$, respectively.
The green curve is the static exchange hole from the Hartree-Fock approximation.
For $t=-4.62$ (solid blue) and $R=2$ 
the exchange hole is virtually indistinguishable from the static one.
}
\label{fig:rhoxcAv}%
\end{center}
\end{figure}

The relevant quantity determining the exchange-correlation potential
is the first radial moment of the spherical average of the
exchange-correlation holes. In the upper panel of Fig. \ref{fig:rhoxcAv} the 
spherical average of the real part of the exchange hole multiplied by $R'$ is shown
for several values of $R$ and $t$.
When taking a spherical average the factor $\sin{\theta}$ implies that
the angular region around $\theta\approx 0$ contributes significantly less than
the region around $\theta\approx \pi/2$. This is confirmed by comparing
Fig. \ref{fig:rhoxcAv}
and Fig. \ref{fig:rhoxR2T4-34} for $R=2$ in which it can be seen that the behavior of the
spherical average essentially follows that of the exchange hole for $\theta=\pi/2$.

For $R=2$ and $t=-4.62$ the exchange hole is virtually indistinguishable from the static
Hartree-Fock exchange hole. As expected, the exchange hole for small $R$ and $t$ should
approach the static exchange hole. 
However, even for large values of $R$ and $t$ the exchange hole
still follows closely the static one, as can be seen in the top panel of Fig. \ref{fig:rhoxcAv}.
As $R$ is increased, some weight is transferred from the small to the large regions of $R'$
and the exchange hole appears to become less dependent on $t$ for large $R$.

The spherical average of the correlation hole for several values of $R$ and $t$ is shown in 
the lower panel of
Fig. \ref{fig:rhoxcAv}. The general behavior of the correlation hole
is similar to that of the exchange hole, except for $R=2$ in which
the correlation hole fluctuates strongly around the origin as a function of time 
than the corresponding exchange hole.
As also expected from the sum rule, which integrates to zero for the correlation hole,
more oscillations can be observed than in the corresponding exchange holes.
For $t>0$ the correlation hole tends to be positive around the origin and vanishes
as $R$ increases.

\begin{figure}[H]
\begin{center} 
\includegraphics[scale=0.6, viewport=3.2cm 8cm 17.8cm 20cm, clip,width=\columnwidth]
{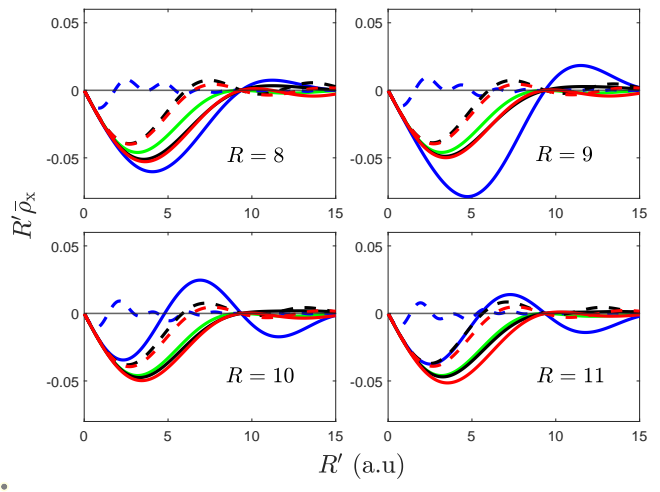}
\hfill
\includegraphics[scale=0.6, viewport=3.2cm 8cm 17.8cm 20cm, clip,width=\columnwidth]
{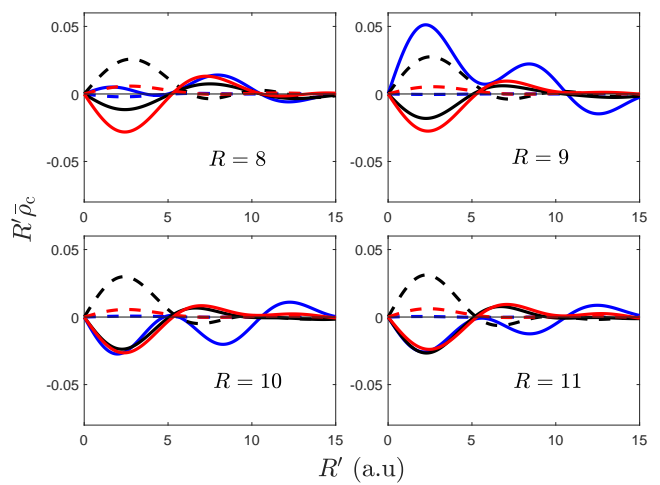}
\includegraphics[scale=0.6, viewport=3.2cm 8cm 17.8cm 20cm, clip,width=\columnwidth]
{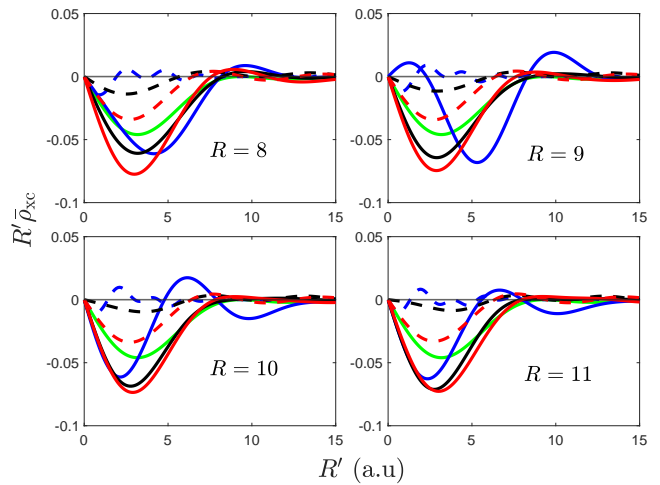}
\caption{
The real part of the spherical average of the 
exchange hole (upper panel), the correlation hole (middle panel), and
the exchange-correlation hole (lower panel)
multiplied by $R'$ for $t=\pm 4.62$ (blue), $\pm 34.75$ (black), $\pm 69.5$ (red),
and $R=8,9,10,11$.
The solid and dashed curves correspond to $t<0$ and $t>0$, respectively.
The green curve is the static exchange hole from the Hartree-Fock approximation.
}
\label{fig:rhoxcAv1}%
\end{center}
\end{figure}

The exchange and the correlation holes exhibit strong fluctuations, deviating greatly
from the static one at $t=- 4.62$, which
is found to correlate with the radial distance. To investigate further, the exchange and
the correlation holes in the vicinity of $R=10$ are calculated and shown in Fig. \ref{fig:rhoxcAv1}.
It is quite evident that the large fluctuations at $R=9-10$ correlate with the time $t=- 4.62$.
These large fluctuations originate from the small values of $|G_0(R,t)|$ at these values of $R$ and $t$,
and are carried over to the exchange-correlation potentials
as discussed in the next section.


\subsection{Exchange-correlation potentials}

\begin{figure}[H]
\begin{center} 
\includegraphics[scale=0.6, viewport=3cm 8cm 17cm 20cm, clip,width=\columnwidth]
{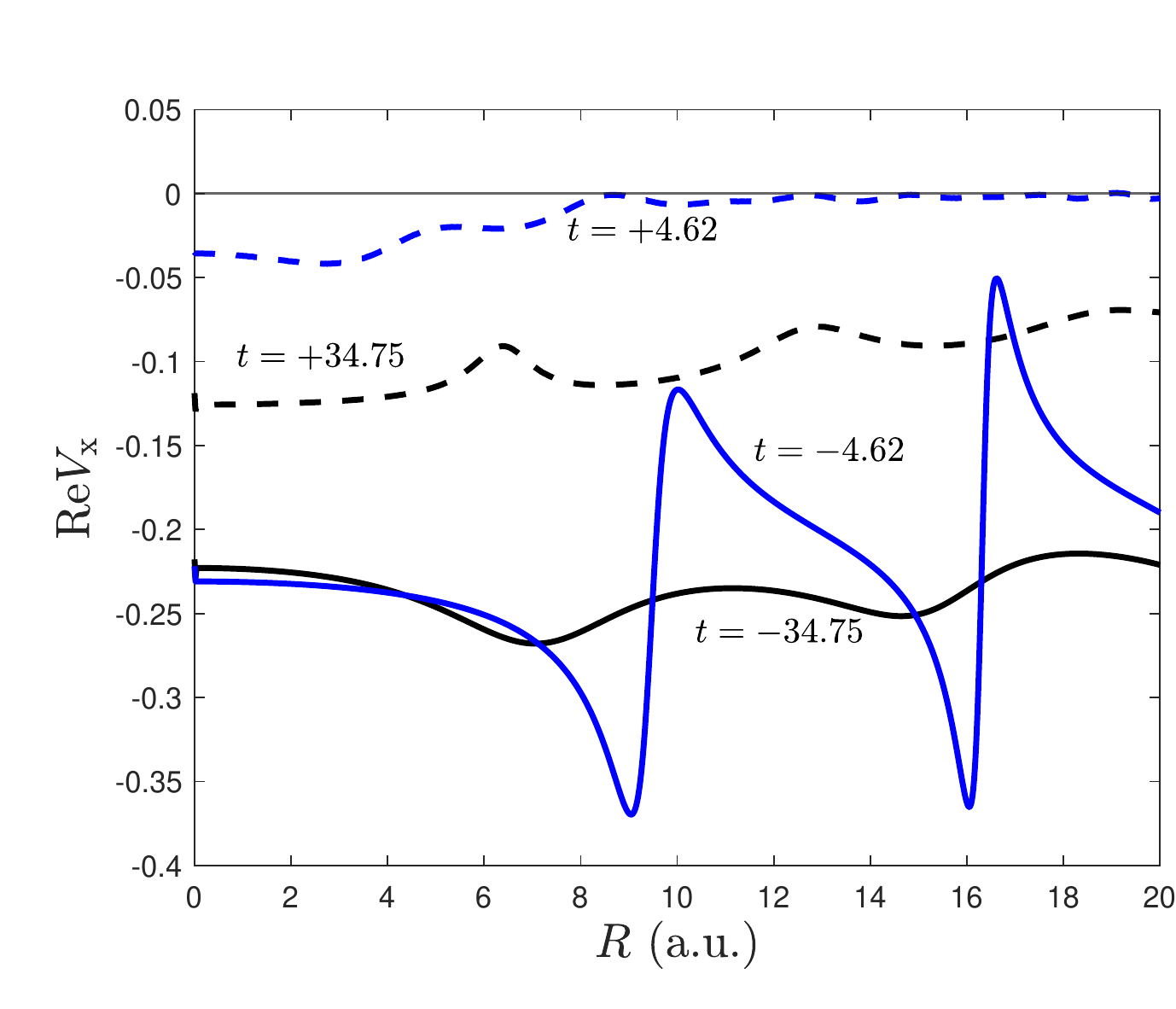}
\hfill
\includegraphics[scale=0.6, viewport=3cm 8cm 17cm 20cm, clip,width=\columnwidth]
{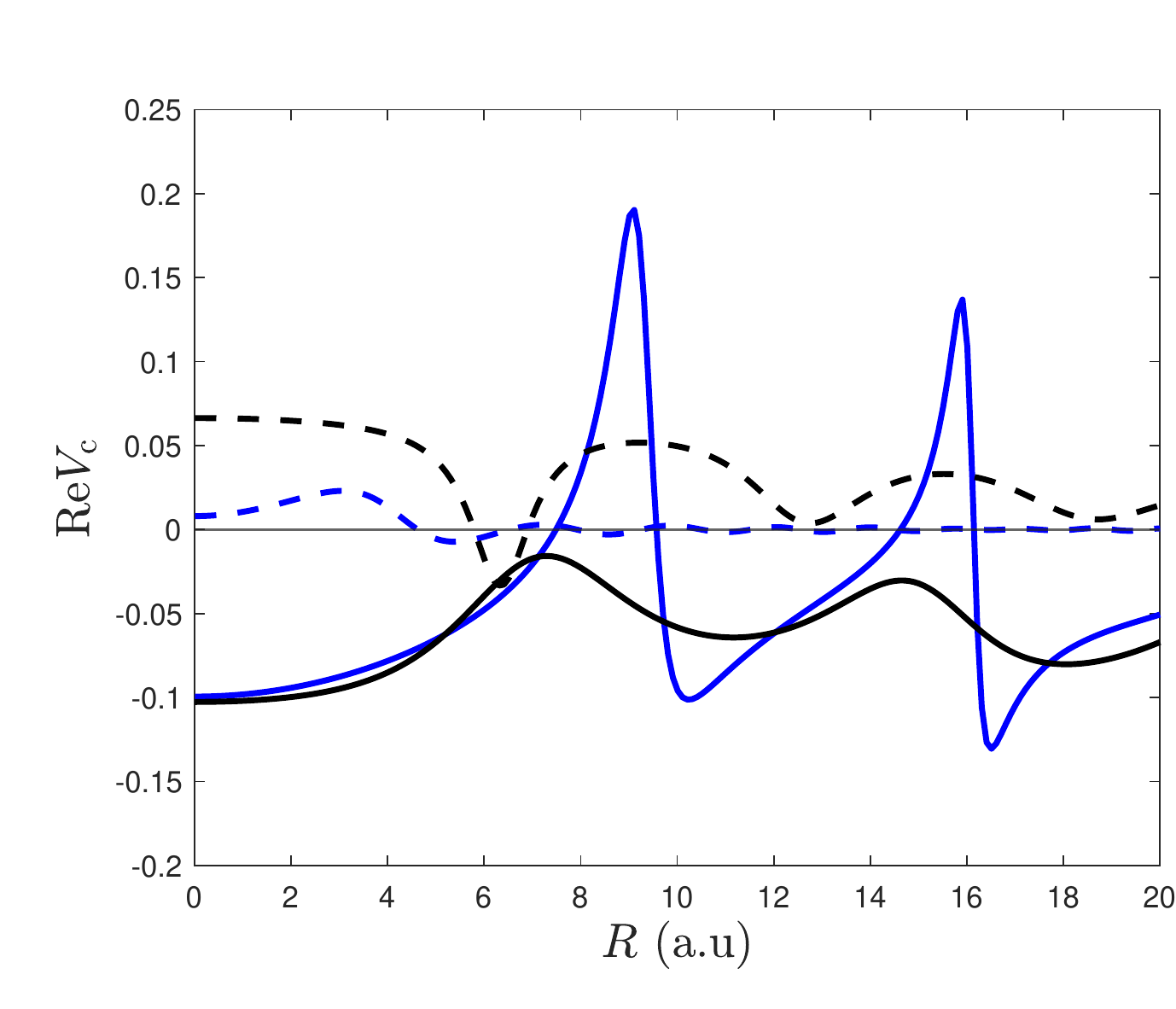}
\hfill
\includegraphics[scale=0.6, viewport=3cm 8cm 17cm 20cm, clip,width=\columnwidth]
{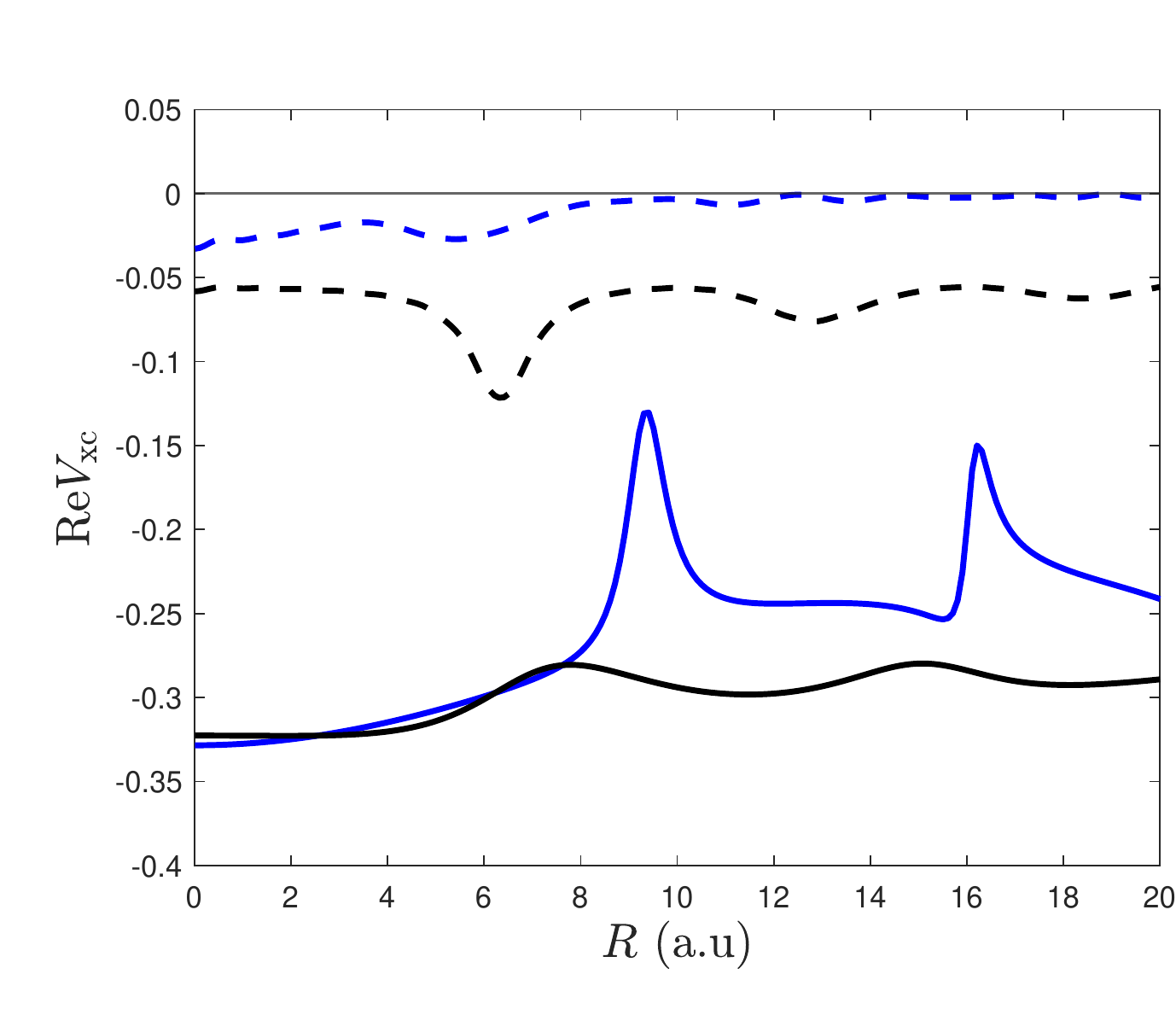}
\caption{
The real part of the exchange potential (top), the correlation potential (middle),
and the exchange-correlation potential for $t=\pm 4.62$ (blue) and $\pm 34.75$
(black). The solid and dashed curves correspond to $t<0$ and $t>0$, respectively.
}
\label{fig:ReVxcT4-34}%
\end{center}
\end{figure}

\begin{figure}[H]
\begin{center} 
\includegraphics[scale=0.6, viewport=3cm 8cm 17cm 20cm, clip,width=\columnwidth]
{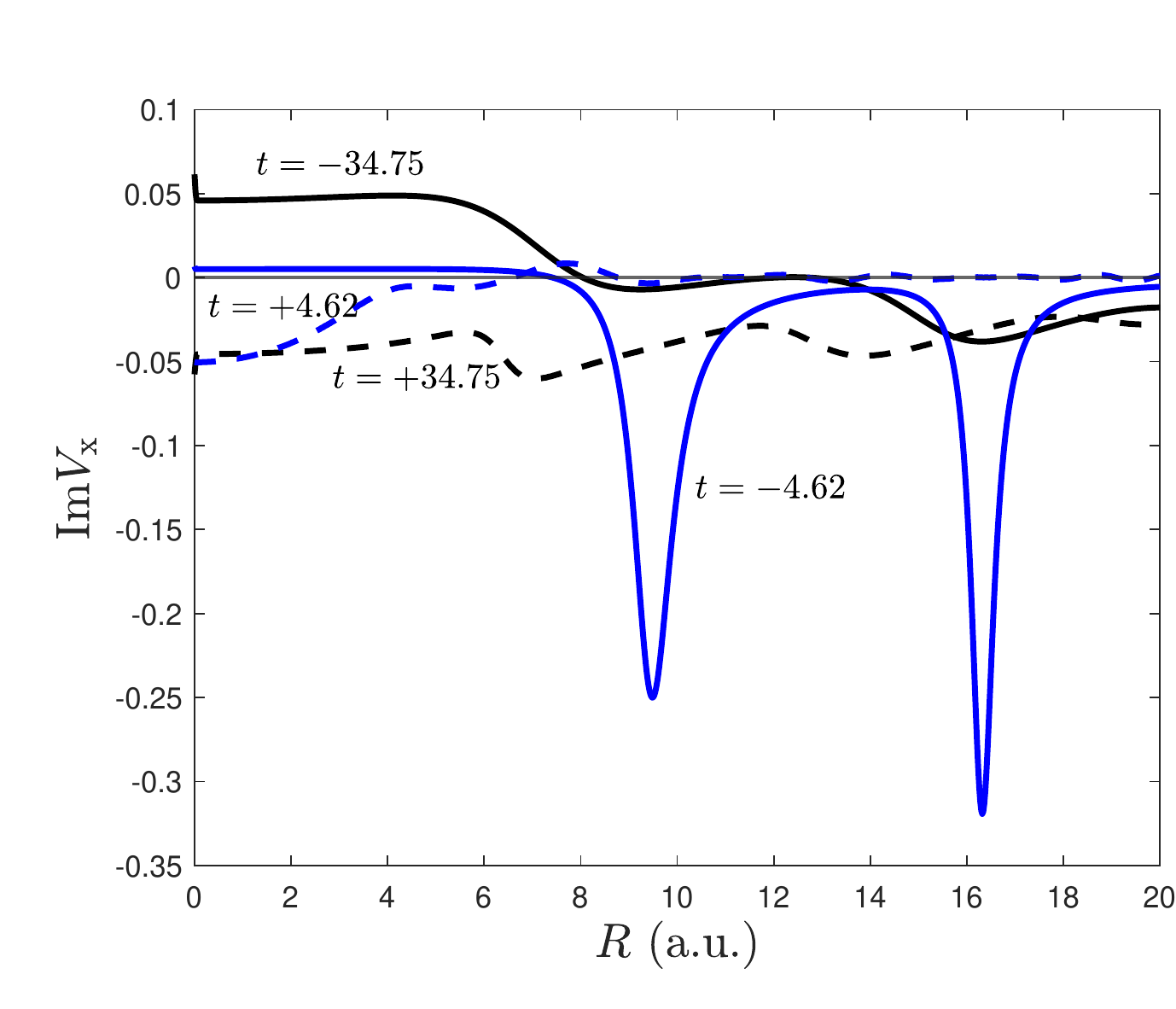}
\hfill
\includegraphics[scale=0.6, viewport=3cm 8cm 17cm 20cm, clip,width=\columnwidth]
{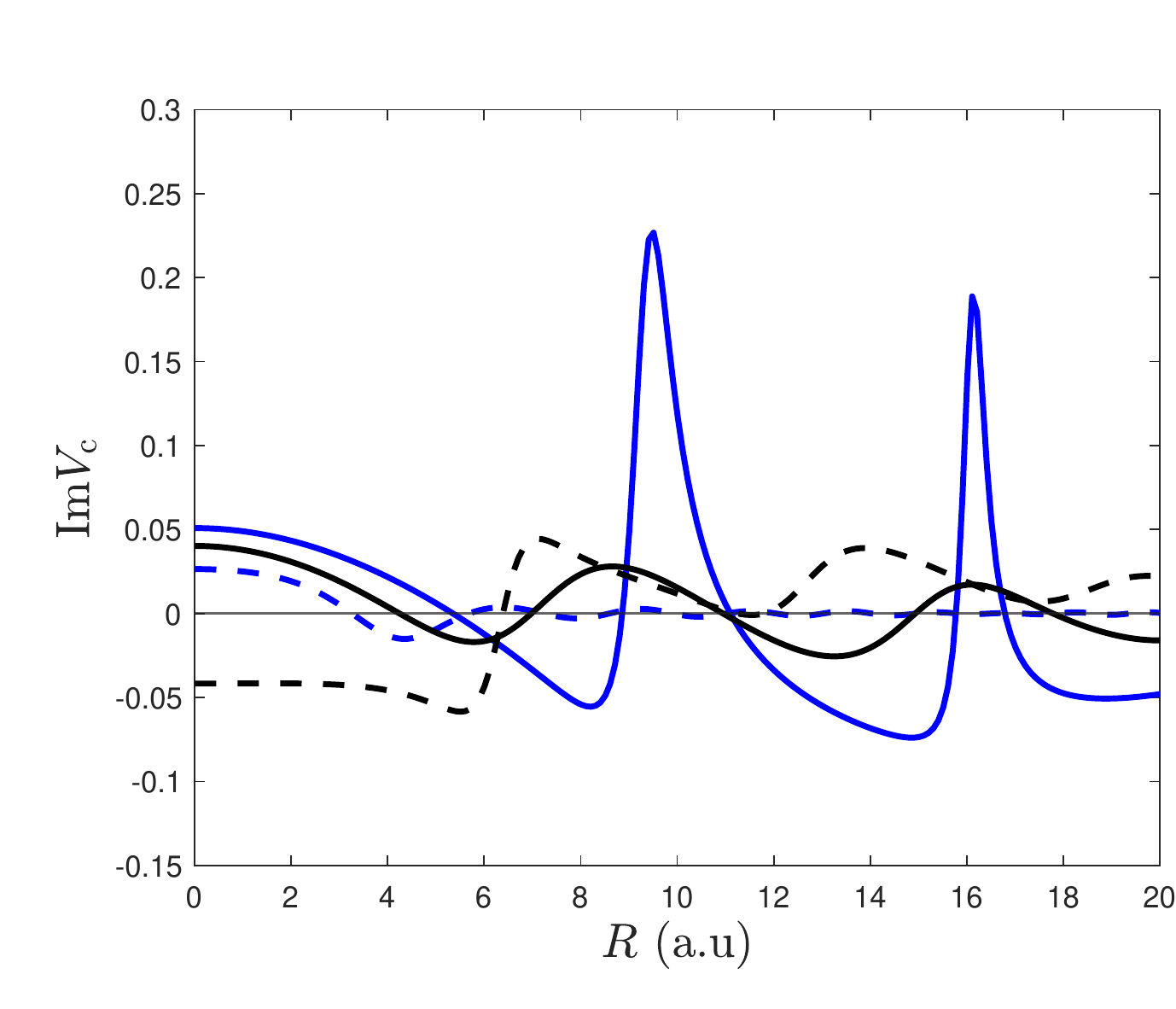}
\hfill
\includegraphics[scale=0.6, viewport=3cm 8cm 17cm 20cm, clip,width=\columnwidth]
{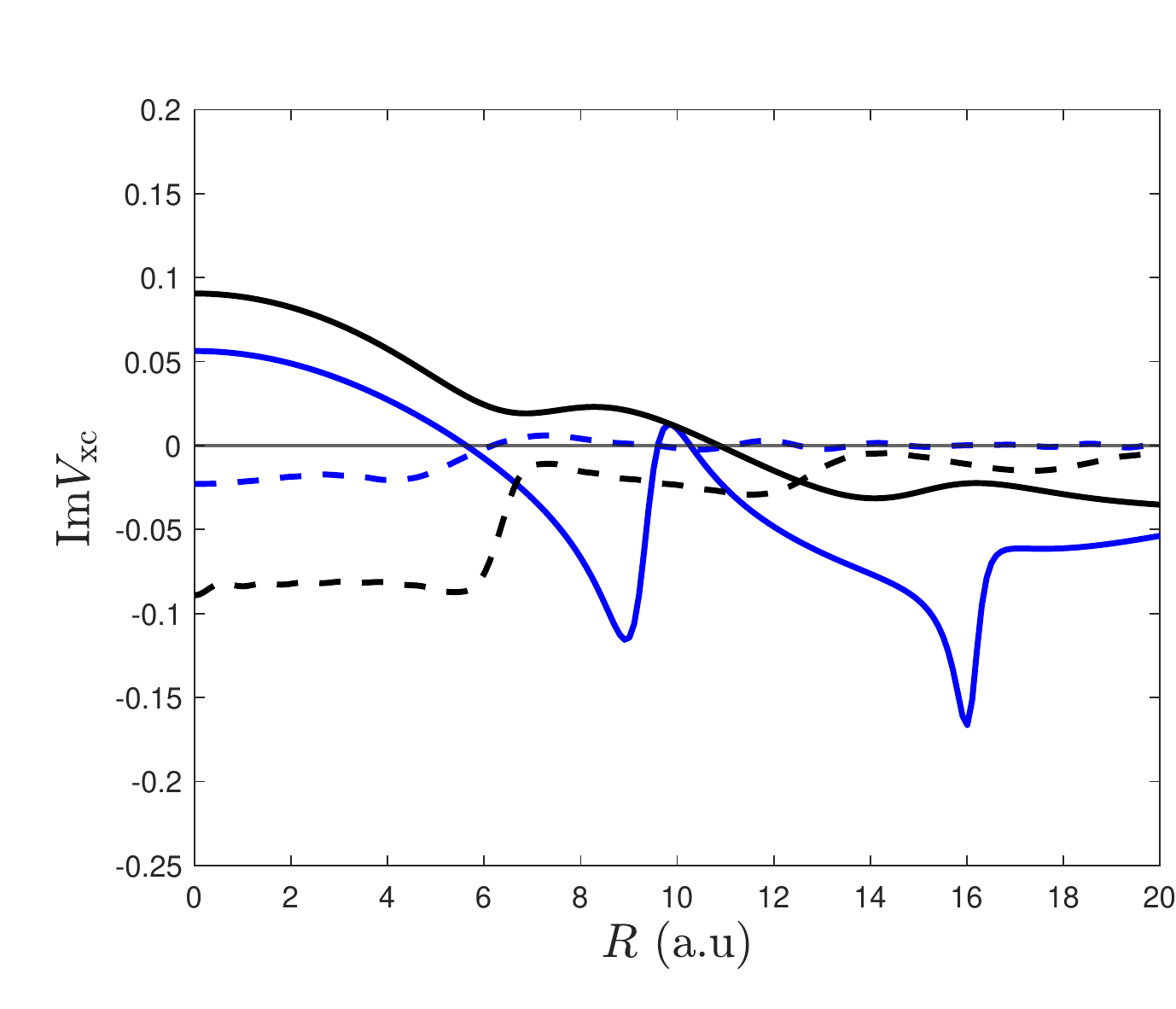}
\caption{
The imaginary part of the exchange potential (top), the correlation potential
(middle), and the exchange-correlation potential (bottom) for $t=\pm 4.62$ (blue) and $\pm 34.75$
(black). The solid and dashed curves correspond to $t<0$ and $t>0$, respectively.
}
\label{fig:ImVxcT4-34}
\end{center}
\end{figure}

\begin{figure}[t]
\begin{center} 
\includegraphics[scale=0.6, viewport=3cm 8cm 17cm 20cm, clip,width=\columnwidth]
{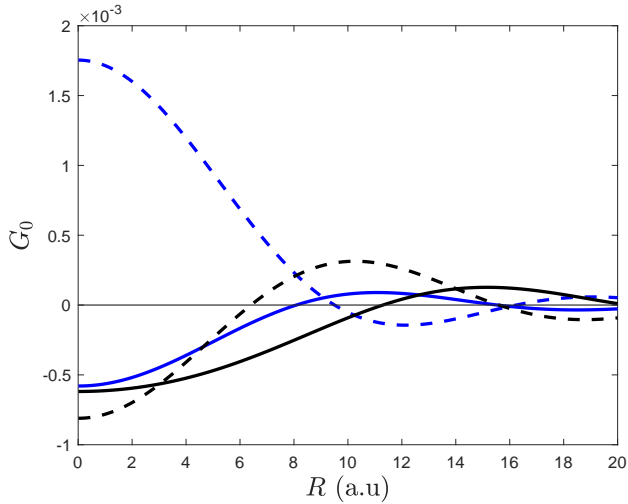}
\caption{
The real (solid) and the imaginary (dashed) parts
of the non-interacting Green function for $t=-4.62$ (blue) and $-34.75$ (black). 
For $t=-4.62$, the value of $|G_0|$ is close to zero at positions $R=9$ and $16$,
whereas for $t=-34.75$, it remains relatively substantial.
}
\label{fig:G0T4-34}%
\end{center}
\end{figure}

\begin{figure}[t]
\begin{center} 
\includegraphics[scale=0.6, viewport=3cm 8cm 17cm 20cm, clip,width=\columnwidth]
{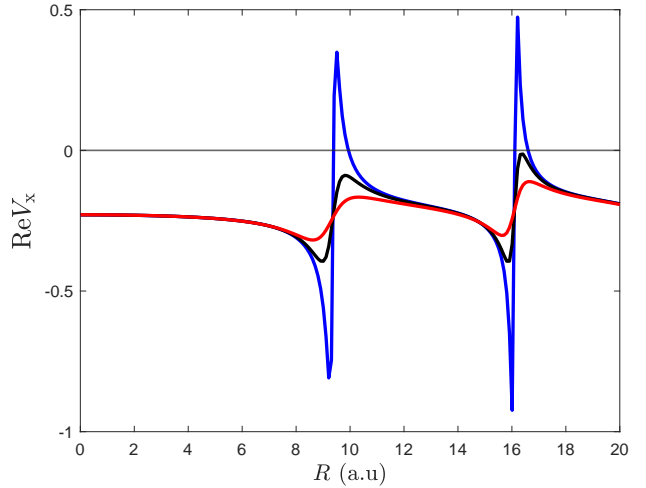}
\caption{
The real part of the exchange potential for $t=-1$ (blue), $-4$ (black)
and $-8$ (red). The structure at $R=9$ and $16$ becomes sharper as $t\rightarrow 0^-$,
which, however, is largely cancelled by the correlation potential, as illustrated in
Fig. \ref{fig:VxcT1}.
}
\label{fig:VxT1-4-8}%
\end{center}
\end{figure}

\begin{figure}[t]
\begin{center} 
\includegraphics[scale=0.9, viewport=6cm 4.2cm 18cm 24cm, clip,width=\columnwidth]
{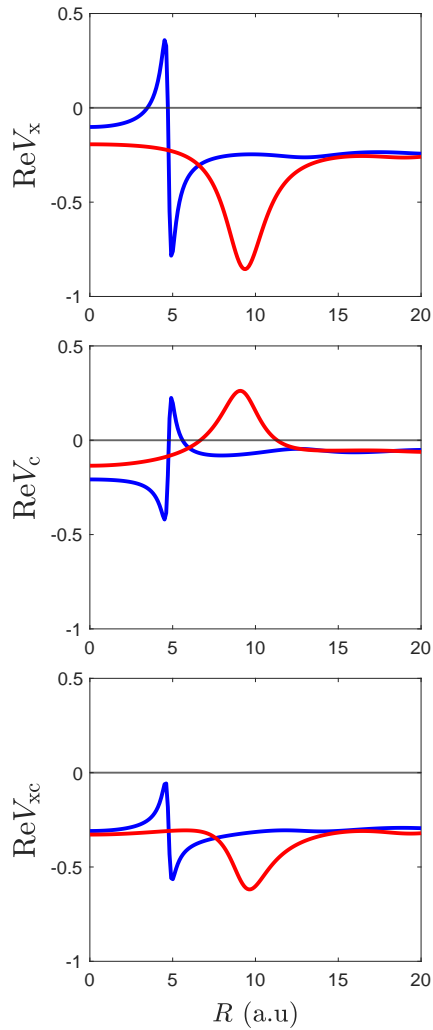}
\caption{
The real part of the exchange potential (top), the correlation potential (middle),
and the exchange-correlation
potential (bottom) for $t=-60$ (blue) and $-84$ (red).
}
\label{fig:VxcT60-84}%
\end{center}
\end{figure}

\begin{figure}[t]
\begin{center} 
\includegraphics[scale=0.6, viewport=3cm 8cm 17cm 20cm, clip,width=\columnwidth]
{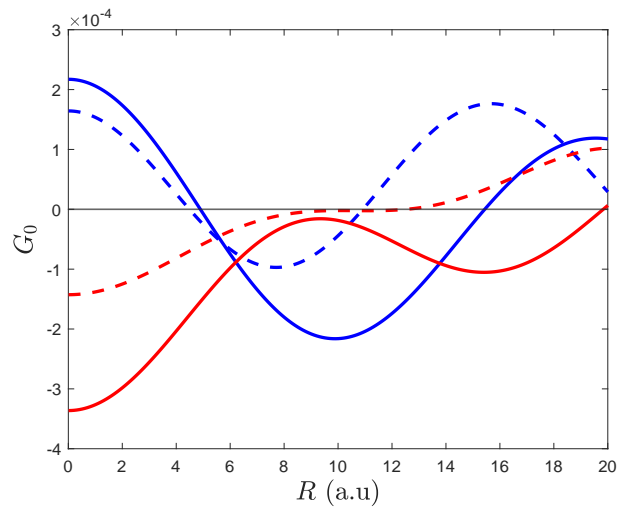}
\caption{
The real (solid) and the imaginary (dashed) parts
of the non-interacting Green function for $t=-60$ (blue) and $-84$ (red). 
}
\label{fig:G0T60-84}%
\end{center}
\end{figure}

From the spherical average of the exchange-correlation hole
the exchange-correlation potential can be determined. 
The real and imaginary parts of the exchange potential for $t=\pm 4.62$ and $\pm 34.75$ are shown in 
Figs. \ref{fig:ReVxcT4-34} and \ref{fig:ImVxcT4-34}, respectively. 
For $t<0$ 
there is a striking difference between the exchange potentials
corresponding to the two times. The exchange
potential corresponding to $t=-4.62$ exhibits a pronounced structure 
around $R=9$ and $16$, whereas the one corresponding to $t=-34.75$
is almost constant with a relatively weak feature. 
As mentioned earlier, this strong structure at $t=-4.62$ arises from the small 
value of $|G_0(R,t)|$ at $R=9$ and $16$ as shown in Fig. \ref{fig:G0T4-34}. 
In fact, the exchange
potential has a strong structure at those positions for a range of $t$,
as can be seen in Fig. \ref{fig:VxT1-4-8}. The structure is very sharp for
very small $t$ and as the magnitude of $t$ increases, it becomes weaker.
A plausible explanation is that when a hole is introduced into the system, large density
fluctuations arise and within a short period of time the system does not have 
enough time to relax, generating as a consequence a strong spatial variation
in the exchange potential.
However, strong spatial variations also appear at a large-time scale as shown in 
Fig. \ref{fig:VxcT60-84}.
The common denominator for the presence of strong structures in the exchange-correlation
potential is the diminishing value of $|G_0(R,t)|$ at the corresponding position $R$ 
and time $t$, as can be seen in Fig. \ref{fig:G0T60-84}.

For $t>0$ the exchange potentials are generally weaker and less attractive
than for $t<0$ as can be seen in Fig. \ref{fig:ReVxcT4-34}.
The difference between the two cases may be traced back to the fundamental 
physical difference between removing
an electron (hole creation, $t<0$) and adding an electron ($t>0$), in that  
the removed electron is part of the system in the ground state whereas the added
electron is not. This is also reflected in the sum rule which integrates to $-1$ for
a hole addition but zero for an electron addition. Thus, the exchange potential
associated with a hole creation must be stronger than that associated with an electron 
addition. Aside from special systems that possess a particle-hole symmetry,
the creation of a hole can be expected to create a stronger disturbance to the system
in the ground state than the addition of an electron.

The exchange potential does not take into account the classical
Coulomb response of the electrons which results in screening of the density fluctuations
arising from the exchange interaction. In Fig. \ref{fig:ReVxcT4-34} (middle) the real parts of the 
correlation
potentials associated with the linear density response of the system are shown for
the two pairs of representative times $t=\pm 4.62$ and $\pm 34.75$.
As in the case of exchange, the correlation potential corresponding to $t=-4.62$
has a pronounced structure in comparison with the one corresponding to $t=-34.75$.

The real and imaginary parts of the
exchange-correlation potentials are displayed in Figs. \ref{fig:ReVxcT4-34}
and \ref{fig:ImVxcT4-34}, respectively.
Peaks in the imaginary part of
the exchange-correlation potential, which is
the analog of
the imaginary part of the self-energy, signals the presence of damping.
It is evident that there is a strong cancellation of exchange and correlation.
The Kramers-Kronig-like 
structures in the exchange potential between $R=8$ and $10$ as well as between
$15$ and $17$ in Fig. \ref{fig:ReVxcT4-34} 
induce polarization in which electrons and holes are accumulated
in opposite regions where the minimum and maximum of $V_\mathrm{x}$ are located.
This exchange polarization is neutralized by Coulomb screening, which is affected
by the presence of inverted structures in $V_\mathrm{c}$ at the corresponding regions.
This results in a strong cancellation between exchange and correlation potentials and the 
resulting total exchange-correlation potential has a much less structure than both the exchange and the
correlation potentials.

\begin{figure}[t]
\begin{center} 
\includegraphics[scale=0.6, viewport=3cm 8cm 17cm 20cm, clip,width=\columnwidth]
{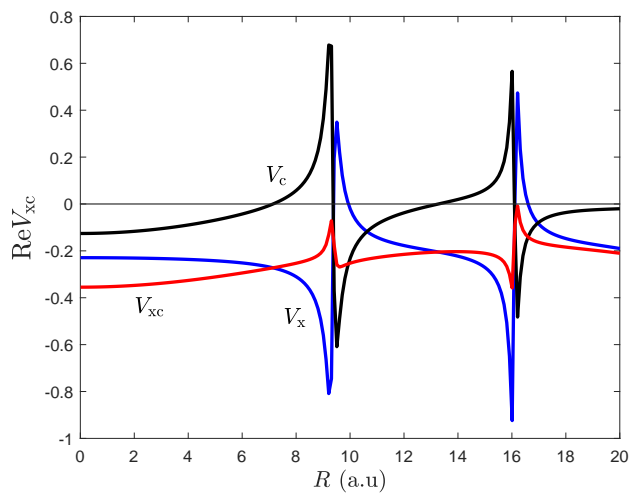}
\caption{
The real part of the exchange potential (blue), the correlation potential (black),
and the exchange-correlation potential (red) for $t=-1$. The large cancellation 
between exchange and correlation can be clearly seen.
}
\label{fig:VxcT1}%
\end{center}
\end{figure}

To illustrate further the large cancellation between exchange and correlation, 
the real part of the exchange, correlation, and exchange-correlation potentials for 
$t=-1$ are
displayed in Fig. \ref{fig:VxcT1}. 
It is tempting to speculate that the remaining structure in
the exchange-correlation potential might be smoothed out when full RPA calculations are
carried out without relying on the plasmon-pole approximation.
A similar cancellation is also found for $t=-34.75$ (Fig. \ref{fig:ReVxcT4-34})
but to a much lesser extent since both the exchange and the correlation potentials
have little structures to begin with. 

The strong cancellation between exchange and correlation is
well known in the self-energy formalism. The cancellation manifests itself
in the one-particle dispersion in which the too large 
occupied bandwidth within the Hartree-Fock approximation
is significantly reduced by correlation
effects, resulting in a dispersion close to the non-interacting one.
The $V_\mathrm{xc}$ formalism, however, offers a different perspective
in which the cancellation can be explicitly seen in the 
exchange-correlation potential in space and time.

\begin{table}[h!]
\centering
\begin{tabular}{||c| c| c| c| c| c||} 
 \hline
 $R$ & 9 & 5 & 9 & 16 & 5\\ [1ex] 
 \hline
$t$ & 0 to -5 & -60 & -84 & -102 & -114\\
 \hline
 \end{tabular}
\caption{
The approximate values of $R\leq 20$ and the corresponding times $t\leq 0$
around which strong structures are found in the exchange and the correlation potentials
and strong deviations in the exchange and the correlation holes from their static 
counterparts are observed.
}
\label{table:1}
\end{table}

%

\subsection{Comparison with the model Green function}

\begin{figure}[t]
\begin{center} 
\includegraphics[scale=0.6, viewport=3cm 8cm 17cm 20cm, clip,width=\columnwidth]
{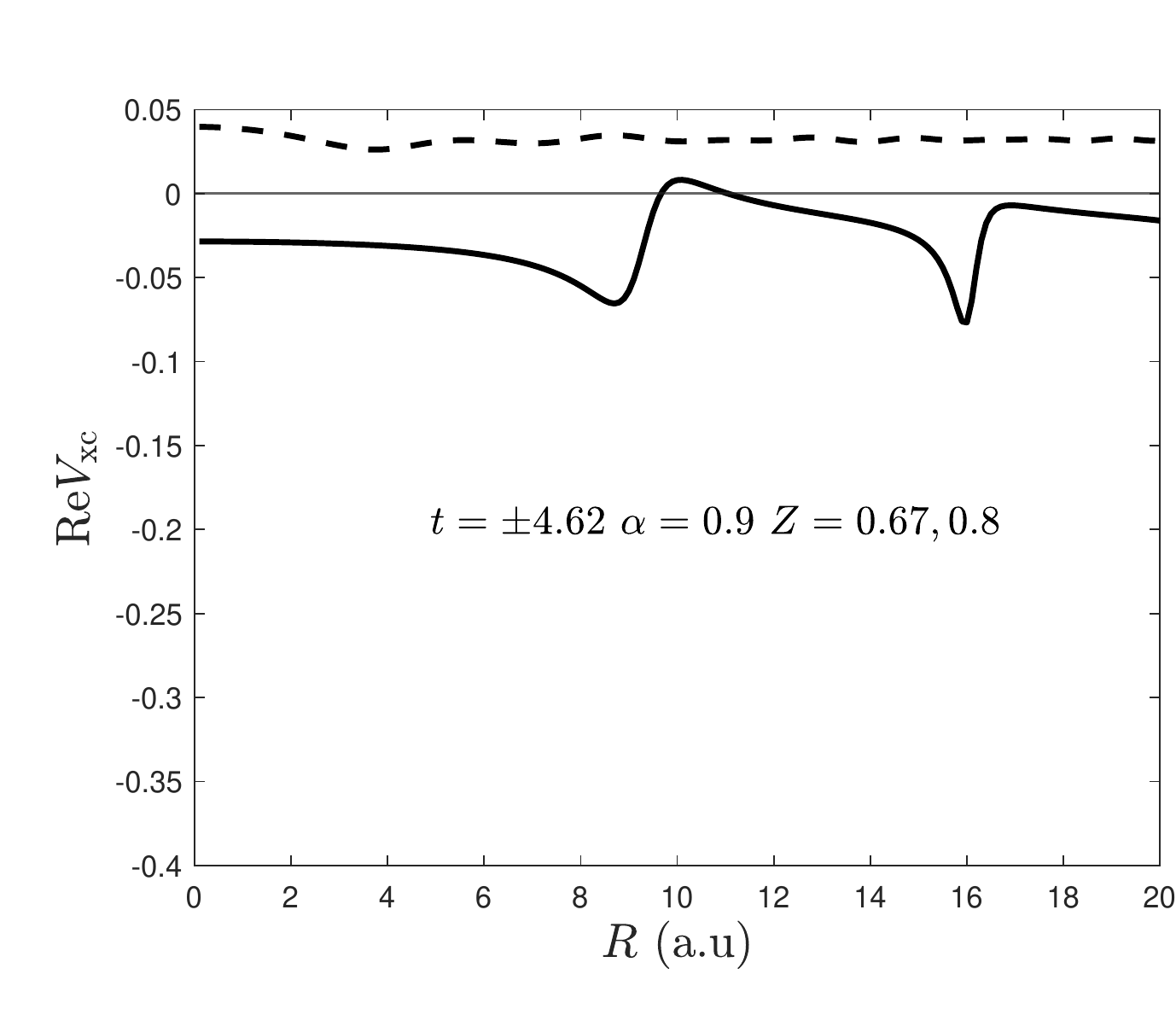}
\hfill
\includegraphics[scale=0.6, viewport=3cm 8cm 17cm 20cm, clip,width=\columnwidth]
{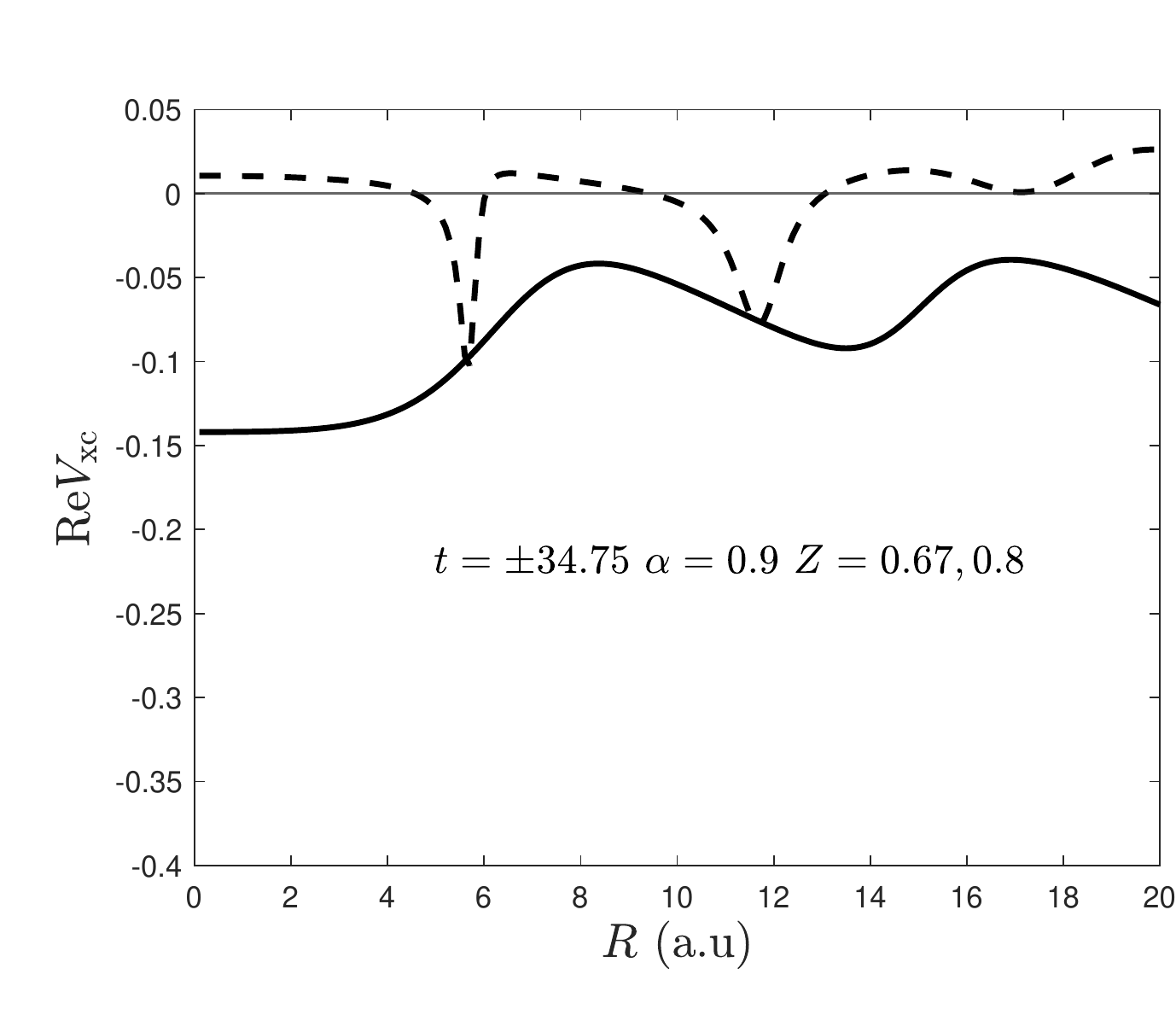}
\end{center}

\caption{
The real part of the 
exchange-correlation potential for $t=\pm 4.62$ (top) and $t=\pm 34.75$ (bottom) from the model Green function.
The solid and dashed curves correspond to $t<0$ and $t>0$, respectively.
The parameters used are indicated in the figure. Two different values of renormalization factors
$Z$ are used, one for $t<0 \,(Z=0.67)$ and one for $t>0 \,(Z_\mathrm{p}=0.8)$. The choice of the
parameters is based on the $GW$ results.
}
\label{fig:VxcModel}
\end{figure}

To check the electron gas results and to gain more insight into the salient features of the
exchange-correlation potentials, a model Green function has been constructed as described in Sec.
\ref{sec:ModelG}.  The exchange-correlation potentials 
derived from the model for
$t=\pm 4.62$ and $t=\pm 34.75$ are shown in Fig. \ref{fig:VxcModel}, 
which should be compared with Fig. \ref{fig:ReVxcT4-34}. Apart from the absolute 
position, which cannot be determined without calculating the quasiparticle dispersion from the
resulting electron gas exchange-correlation potential, the agreement for $t=\pm 34.75$
is quite striking. There is a discrepancy in the separation between $V_\mathrm{xc}(t>0)$ and
$V_\mathrm{xc}(t<0)$, which may indicate an overestimation of correlation within the plasmon-pole
approximation. Increasing the range of integration from $2k_\mathrm{F}$
to $4k_\mathrm{F}$ in the integration over the unoccupied states results in a down shift of the
correlation potential while the shape remains stable. 
There is less agreement for $t=\pm 4.62$, both concerning the shape 
as well as the separation between $V_\mathrm{xc}(t>0)$ and
$V_\mathrm{xc}(t<0)$. Again, this may be due to the plasmon-pole approximation, or,
speculatively speaking, to the model itself, which may not be accurate for small $t$.
The small time behavior is determined by the quality of the model at high energy.
The model assumes a sharp plasmon excitation, which may be valid for small momentum 
but becomes less accurate at large momenta beyond the critical momentum.

\begin{figure}[t]
\begin{center} 
\includegraphics[scale=0.6, viewport=3cm 8cm 17cm 20cm, clip,width=\columnwidth]
{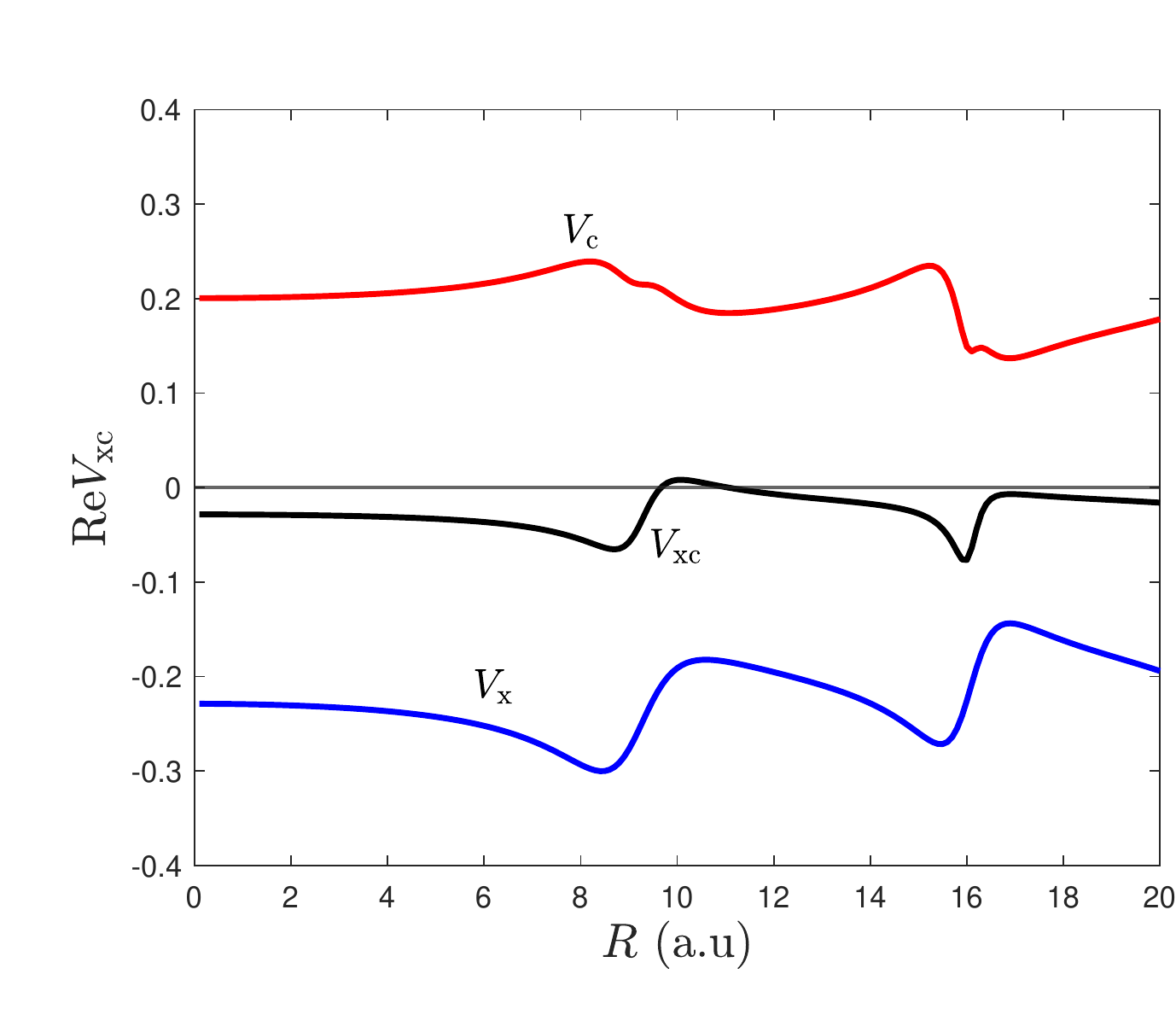}
\hfill
\includegraphics[scale=0.6, viewport=3cm 8cm 17cm 20cm, clip,width=\columnwidth]
{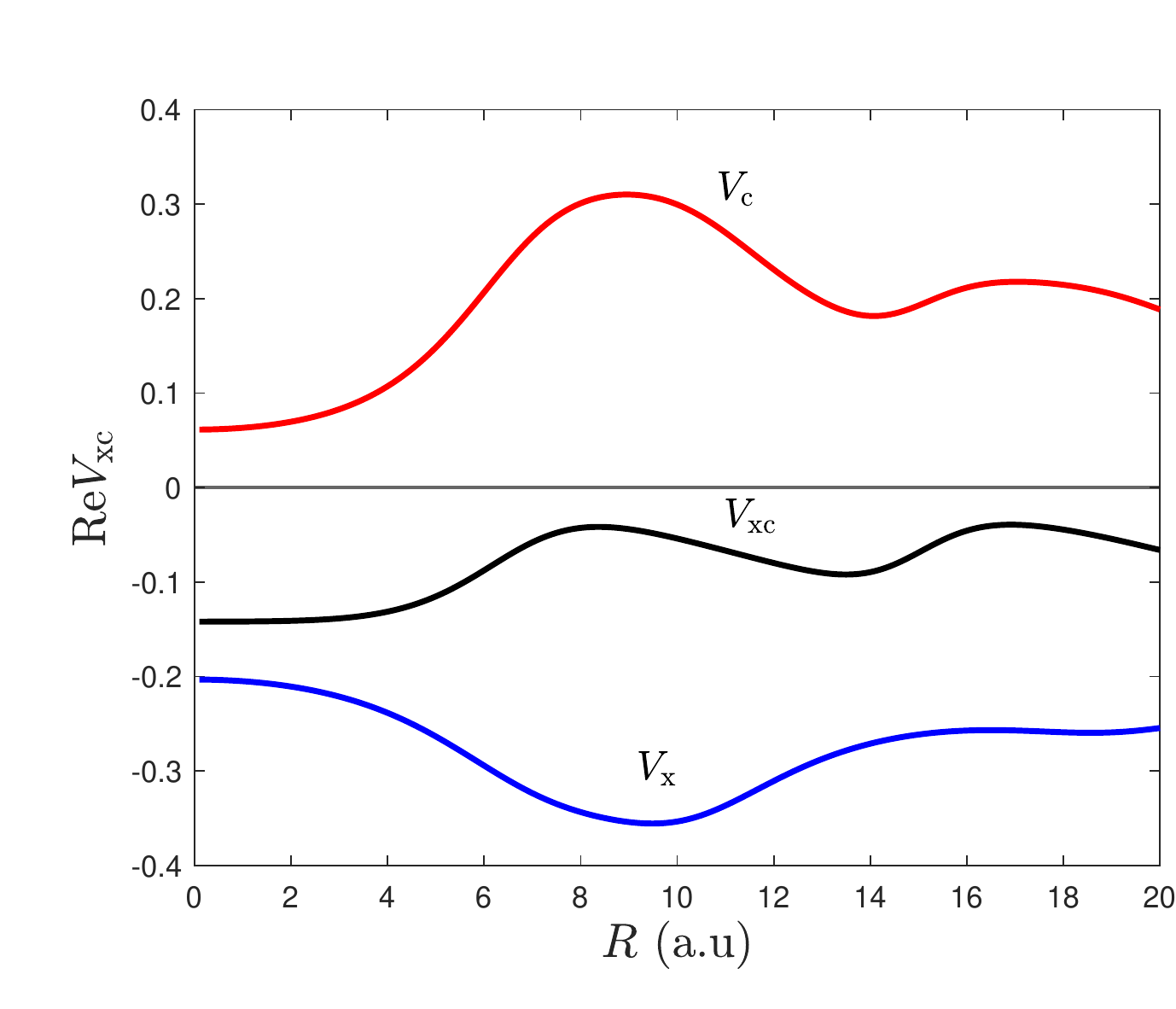}
\end{center}

\caption{
The real part of the exchange potential $V_\mathrm{x}$ (blue)
derived from the Hartree-Fock Green function and the 
exchange-correlation potential $V_\mathrm{xc}$ (black) from the model Green function for $t=- 4.62$
(top) and $t=- 34.75$ (bottom).
The correlation potential (red) is defined as $V_\mathrm{c}=V_\mathrm{xc}-V_\mathrm{x}$.
}
\label{fig:VxcModelFull}
\end{figure}

To make a separate comparison of both $V_\mathrm{x}$ and $V_\mathrm{c}$, 
the exchange potential
deduced from the Hartree-Fock Green function is calculated and used to determine
the correlation potential of the model by subtracting the exchange potential from the
total exchange-correlation potential. The results are shown in Fig. \ref{fig:VxcModelFull}
for $t=\pm 4.62$ and $t=\pm 34.75$.
Here, there is no ambiguity regarding the absolute position of the exchange 
potentials.
The close agreement between the exchange potential deduced from the Hartree-Fock Green function and
the one calculated from Eqs. (\ref{eq:Vx-}) and (\ref{eq:Vx+}) for the electron gas attests
that
the physical interpretation of the exchange-correlation hole in Eq. (\ref{eq:xc-hole}), in which the first term on the 
right-hand side is associated with the exchange hole, is well motivated.


\section{Conclusion and summary}

The exchange-correlation hole and potential of the homogeneous electron gas
have been studied within the RPA for $r_s=4$ and several
representative time periods. 
The angular dependence of the exchange-correlation hole for $t<0$ shows a
stronger oscillation along the line joining the location where the hole is created and
the location where the hole is annihilated ($\theta=0$) whereas along 
the perpendicular
direction ($\theta=\pi/2$) it shows the least oscillation. 
This behavior can be attributed to
the degree of preponderance of the hole along the respective directions.

The behavior of the exchange and the correlation holes reveals a correlation between the separation $R$
and the time period $t$. For certain values of $R$ large fluctuations are found within
a range of time. This behaviour is mimicked by the exchange and the correlation potentials and
is found to originate from the diminishing value of $|G_0(R,t)|$ at the respective position and time.

The spherical average of the exchange-correlation hole for $t<0$ is generally larger than for $t>0$,
which is a consequence of the sum rule being $-1$ for the former and $0$ for the latter.
This property is inherited by the exchange-correlation potential, since it is determined by the first radial
moment of the spherical average of the exchange-correlation hole. It is found that the exchange
potential in space and time
is substantially cancelled by the correlation potential, which maybe seen
as the analog of the well-known cancellation of the Fock exchange and 
the correlation self-energy
of the electron gas in momentum and frequency. The strong cancellation results in
an exchange-correlation potential with much less structure than in the corresponding exchange and
correlation potentials. This encouraging result lends support for the feasibility of 
applying the local density approximation.

Analysis of the sum rule offers a physical explanation why using
a non-interacting Green function is more advantageous than using a renormalized one when
calculating the response function in RPA and consequently in calculating the self-energy
within the $GW$ approximation.
A simple vertex correction that preserves the sum rule is then proposed.

The present work provides a starting point for more accurate calculations of
the exchange-correlation hole and potential of the electron gas.
The plasmon-pole approximation is not expected to yield accurate results since it neglects
the limits in the solid angles. Nevertheless, comparison with results obtained from a model Green
function suggests that it captures the main features of the exchange-correlation potential.
It would be highly desirable to perform full RPA calculations without employing the plasmon-pole
approximation with the aim of constructing
a local-density type approximation for the exchange-correlation potential, which can be applied to
calculate the Green function of real materials, circumventing computationally
expensive traditional self-energy calculations based on Feynman diagrams and
path integral techniques.

\begin{acknowledgments}
Financial support from the Knut and Alice Wallenberg (KAW) 
Foundation (Grant number 2017.0061)
and the Swedish Research Council (Vetenskapsrådet, VR, Grant number 2021\_04498) 
is gratefully acknowledged. We thank Rex Godby for valuable discussions.
\end{acknowledgments}

\appendix

\section{Calculation of $G_0(R,t>0)$}
\label{app:G0}

The integral over $k$ in $G_0(R,t>0)$ can be decomposed as follows:
\begin{align}
\int_{k_\mathrm{F}}^\infty dk
=\int_{0}^\infty dk - \int_0^{k_\mathrm{F}} dk.
\end{align}
\begin{equation}
iG_0(R,t>0)=\frac{1}{2\pi^{2}R}\left\{  I(0,\infty)-I(0,k_{\text{F}})\right\}  ,
\end{equation}
where%
\begin{equation}
I(a,b)=\int_{a}^{b}dk\text{ }k\sin(kR)e^{-ik^{2}t/2}.
\end{equation}

Consider the following integral
\begin{align}
\int_{-\infty}^\infty dk\, k e^{ikR} e^{-ik^2 t/2}
    =e^{iR^2/2t}
    \int_{-\infty}^\infty dk\, k e^{-i(k-R/t)^2 t/2}.
\end{align}
By making a change of variable,
\begin{equation}
    k-\frac{R}{t}= \sqrt{\frac{2}{t}} x,
\end{equation}
the integral $I(0,\infty)$ can be performed analytically yielding
\begin{align}
    I(0,\infty)
    &=\sqrt{\frac{\pi}{2it}}\frac{R}{it} e^{iR^2/2t}.
\end{align}

For $t\rightarrow 0^+$, by introducing a converging factor $e^{-|\alpha|q}$ one finds,
\begin{equation}
\lim_{\alpha\rightarrow{0}} \int_{0}^{\infty}dk\text{ }k\sin(kR)e^{-|\alpha|q}=0 
\end{equation}
implying that indeed $G_0(R,0^+)=G_0(R,0^-)$ for $R\neq 0$.

\section{Correlation hole}
\label{app:rhoc}

From Eq. (\ref{eq:xc-hole}), the correlation hole is given by the following equation:
\begin{align}
    &\rho_\mathrm{c}(r,r',r'';t)G(r,r';t)
    =    \nonumber\\
    & i\int dr_4dt_4\,G(r,r_4;t-t_4) K(r_4,r'';t_4-t)G(r_4,r';t_4).
    \label{eq:c-hole}
\end{align}

A non-interacting $G$ is used,
\begin{align}
    G_0(r-r_4,t-t_4)&=\frac{i}{\Omega}\sum_{k\leq k_F} 
    e^{i\mathbf{k}\cdot(\mathbf{r}-\mathbf{r}_4)} e^{-i\varepsilon_k (t-t_4)}\theta(t_4-t)
    \nonumber\\
    &-\frac{i}{\Omega}\sum_{k> k_F} 
    e^{i\mathbf{k}\cdot(\mathbf{r}-\mathbf{r}_4)} e^{-i\varepsilon_k (t-t_4)}\theta(t-t_4),
\end{align}
\begin{align}
    G_0(r_4-r',t_4)&=\frac{i}{\Omega}\sum_{k'\leq k_F} 
    e^{i\mathbf{k}'\cdot(\mathbf{r}_4-\mathbf{r}')} e^{-i\varepsilon_{k'}
    t_4}\theta(-t_4)
    \nonumber\\
    &-\frac{i}{\Omega}\sum_{k'> k_F} 
    e^{i\mathbf{k}'\cdot(\mathbf{r}_4-\mathbf{r}')} e^{-i\varepsilon_{k'}
    t_4}\theta(t_4),
\end{align}
and
\begin{align}
    &K(r_4-r'',t_4-t)
    \nonumber\\
    &=\frac{1}{\Omega}\sum_q\int \frac{d\omega}{2\pi}\,
    e^{-i\mathbf{q}\cdot(\mathbf{r}_4-\mathbf{r}'')}e^{-i\omega(t_4-t)}K(q,\omega).
\end{align}

Consider the right-hand side of Eq. (\ref{eq:c-hole}) and the case $t<0$. 
From the product of the two Green functions
one obtains four terms but only two survive. The first non-zero term is 
(note the additional factor of $i$ from $iGKG$)
\begin{align}
    A_1&=-\frac{i^3}{\Omega^3}\sum_{k\leq k_F}\sum_{k'> k_F}
    e^{i\mathbf{k}\cdot\mathbf{r}}e^{-i\mathbf{k}'\cdot\mathbf{r}'}
         \nonumber\\
    &\;\;\times\sum_q e^{i\mathbf{q}\cdot\mathbf{r}''}
     \int \frac{d\omega}{2\pi}\,e^{i(\omega-\varepsilon_k)t} K(q,\omega)
    \nonumber\\
    &\;\;\times\int dr_4\, e^{-i(\mathbf{k}-\mathbf{k}'+\mathbf{q})\cdot\mathbf{r}_4}
    \int_0^\infty dt_4\,e^{-i(\omega-\varepsilon_k+\varepsilon_{k'}-i\eta)t_4}
    \nonumber\\
    &=\frac{1}{\Omega^2}\sum_{k\leq k_F}\sum_{k'> k_F}
    e^{i\mathbf{k}\cdot\mathbf{r}}e^{-i\mathbf{k}'\cdot\mathbf{r}'}
    e^{i(\mathbf{k}'-\mathbf{k})\cdot\mathbf{r}''}
         \nonumber\\
    &\;\;\times
    \int \frac{d\omega}{2\pi}\, K(|\mathbf{k}'-\mathbf{k}|,\omega)
    \frac{e^{i(\omega-\varepsilon_k)t}}
    {\omega-\varepsilon_k+\varepsilon_{k'}-i\eta},
    \label{eq:A1a}
\end{align}
which can be rewritten in terms of the radial variables:
\begin{align}
    A_1
    &=\frac{1}{\Omega^2}\sum_{k\leq k_F} e^{-i\mathbf{k}\cdot\mathbf{R}'}
    \sum_{k'> k_F}e^{i\mathbf{k}'\cdot\mathbf{R}''}
         \nonumber\\
    &\;\;\times 
     \int \frac{d\omega}{2\pi}\, K(|\mathbf{k}'-\mathbf{k}|,\omega)
    \frac{e^{i(\omega-\varepsilon_k)t}}
    {\omega-\varepsilon_k+\varepsilon_{k'}-i\eta}.
    \label{eq:A1b}
\end{align}
The second non-zero term is similar to the first and given by
%
%
\begin{align}
    A_2
    &=-\frac{1}{\Omega^2}\sum_{k>k_F}
    e^{-i\mathbf{k}\cdot\mathbf{R}'}
    \sum_{k'\leq k_F}
    e^{i\mathbf{k}'\cdot\mathbf{R}''}
         \nonumber\\
    &\;\;\times 
     \int \frac{d\omega}{2\pi}\, K(|\mathbf{k}'-\mathbf{k}|,\omega)
    \frac{e^{-i\varepsilon_{k'}t}}
    {\omega-\varepsilon_k+\varepsilon_{k'}+i\eta}.
\end{align}

For the case $t>0$ similar consideration yields
\begin{align}
    B_1
    &=\frac{1}{\Omega^2}\sum_{k\leq k_F}
    e^{-i\mathbf{k}\cdot\mathbf{R}'}
    \sum_{k'> k_F}
    e^{i\mathbf{k}'\cdot\mathbf{R}''}
         \nonumber\\
    &\;\;\times 
     \int \frac{d\omega}{2\pi}\, K(|\mathbf{k}'-\mathbf{k}|,\omega)
    \frac{e^{-i\varepsilon_{k'}t}}
    {\omega-\varepsilon_k+\varepsilon_{k'}-i\eta}
\end{align}
\begin{align}
    B_2
    &=-\frac{1}{\Omega^2}\sum_{k> k_F} e^{-i\mathbf{k}\cdot\mathbf{R}'}
    \sum_{k'\leq k_F}e^{i\mathbf{k}'\cdot\mathbf{R}''}
         \nonumber\\
    &\;\;\times 
     \int \frac{d\omega}{2\pi}\, K(|\mathbf{k}'-\mathbf{k}|,\omega)
    \frac{e^{i(\omega-\varepsilon_k)t}}
    {\omega-\varepsilon_k+\varepsilon_{k'}+i\eta}.
    \label{eq:B2b}
\end{align}

To calculate the integral over $\omega$ one utilizes the spectral representation of $K$:
\begin{align}
    K(k,\omega)=\int_{-\infty}^0 d\omega'\,\frac{L(k,\omega')}{\omega-\omega'-i\delta}
    +\int_0^{\infty}d\omega'\,\frac{L(k,\omega')}{\omega-\omega'+i\delta},
\end{align}
where
\begin{equation}
    L(k,\omega)=-\frac{1}{\pi}\mathrm{sign}(\omega) \mathrm{Im} K(k,\omega).
    \label{eq:LkwA}
\end{equation}
The spectral function $L(k,\omega)$ is an odd function in $\omega$.

For the case
$t<0$ the contour integral for $A_1$ in the complex $\omega$ plane is closed in 
the lower-half plane yielding
\begin{align}
&\int \frac{d\omega}{2\pi}\, \frac{1}{\omega-\omega'+i\delta}\times
    \frac{e^{i(\omega-\varepsilon_k)t}}
    {\omega-\varepsilon_k+\varepsilon_{k'}-i\eta}
    \nonumber\\
    &=\frac{-ie^{i(\omega'-\varepsilon_k-i\delta)t}}
    {\omega'-\varepsilon_k+\varepsilon_{k'}-i\eta}.
\end{align}
One obtains, using $L(k,-\omega)=-L(k,\omega)$,
\begin{align}
    A_1
    &=\frac{1}{\Omega^2}\sum_{k\leq k_F}
    e^{-i\mathbf{k}\cdot\mathbf{R}'}
    \sum_{k'> k_F}
    e^{i\mathbf{k}'\cdot\mathbf{R}''}
    e^{-i\varepsilon_{k}t}
         \nonumber\\
    &\;\;\times 
     \int_0^\infty d\omega'\, L(|\mathbf{k}'-\mathbf{k}|,\omega')
    \frac{-ie^{i\omega't}}
    {\omega'+\varepsilon_{k'}-\varepsilon_k}
\end{align}
and
\begin{align}
    A_2
    &=\frac{1}{\Omega^2}\sum_{k>k_F}
    e^{-i\mathbf{k}\cdot\mathbf{R}'}
    \sum_{k'\leq k_F}
    e^{i\mathbf{k}'\cdot\mathbf{R}''}
    e^{-i\varepsilon_{k'}t}
         \nonumber\\
    &\;\;\times 
     \int_0^{\infty} d\omega'\, L(|\mathbf{k}'-\mathbf{k}|,\omega')
    \frac{-i}
    {\omega'+\varepsilon_k-\varepsilon_{k'}}.
\end{align}
For the case $t>0$ one obtains
\begin{align}
    B_1
    &=\frac{1}{\Omega^2}\sum_{k\leq k_F}
    e^{-i\mathbf{k}\cdot\mathbf{R}'}
    \sum_{k'> k_F}
    e^{i\mathbf{k}'\cdot\mathbf{R}''}e^{-i\varepsilon_{k'}t}
         \nonumber\\
    &\;\;\times 
     \int_0^\infty d\omega'\, L(|\mathbf{k}'-\mathbf{k}|,\omega')
    \frac{-i}
    {\omega'+\varepsilon_{k'}-\varepsilon_k}
\end{align}
\begin{align}
    B_2
    &=\frac{1}{\Omega^2}\sum_{k> k_F} e^{-i\mathbf{k}\cdot\mathbf{R}'}
    \sum_{k'\leq k_F}e^{i\mathbf{k}'\cdot\mathbf{R}''}e^{-i\varepsilon_{k}t}
         \nonumber\\
    &\;\;\times 
     \int_0^{\infty} d\omega'\, L(|\mathbf{k}'-\mathbf{k}|,\omega')
    \frac{-ie^{-i\omega't}}
    {\omega'+\varepsilon_k-\varepsilon_{k'}}.
\end{align}

For each $t$ the integral over $\omega'$ can be parametrized as follows:
\begin{align}
    M(q,\omega,t) = 
    \int_0^\infty d\omega'\, L(q,\omega')
    \frac{-ie^{i\omega't}}
    {\omega'+\omega},
    \label{eq:MqwtA}
\end{align}
so that
\begin{align}
    A_1
    &=\frac{1}{\Omega^2}\sum_{k\leq k_F}    
    e^{-i\mathbf{k}\cdot\mathbf{R}'} e^{-i\varepsilon_{k}t}
    \sum_{k'> k_F}
    e^{i\mathbf{k}'\cdot\mathbf{R}''}
    \nonumber\\
    &\times M(|\mathbf{k}'-\mathbf{k}|,\varepsilon_{k'}-\varepsilon_{k},t);
    \label{eq:A1M}
\end{align}
\begin{align}
    A_2
    &=\frac{1}{\Omega^2}\sum_{k>k_F}
    e^{-i\mathbf{k}\cdot\mathbf{R}'}
    \sum_{k'\leq k_F}
        e^{i\mathbf{k}'\cdot\mathbf{R}''}
    e^{-i\varepsilon_{k'}t}
         \nonumber\\
    &\;\;\times 
    M(|\mathbf{k}'-\mathbf{k}|,\varepsilon_{k}-\varepsilon_{k'},0);
    \label{eq:A2M}
\end{align}
\begin{align}
    B_1
    &=\frac{1}{\Omega^2}\sum_{k\leq k_F}
    e^{-i\mathbf{k}\cdot\mathbf{R}'}
    \sum_{k'> k_F}
    e^{i\mathbf{k}'\cdot\mathbf{R}''}e^{-i\varepsilon_{k'}t}
         \nonumber\\
    &\;\;\times 
     M(|\mathbf{k}'-\mathbf{k}|,\varepsilon_{k'}-\varepsilon_{k},0);
     \label{eq:B1M}
\end{align}
\begin{align}
    B_2
    &=\frac{1}{\Omega^2}\sum_{k>k_F}
    e^{-i\mathbf{k}\cdot\mathbf{R}'}
    \sum_{k'\leq k_F}
        e^{i\mathbf{k}'\cdot\mathbf{R}''}
    e^{-i\varepsilon_{k}t}
         \nonumber\\
    &\;\;\times 
    M(|\mathbf{k}'-\mathbf{k}|,\varepsilon_{k}-\varepsilon_{k'},-t).
    \label{eq:B2M}
\end{align}
By defining
\begin{align}
    &\gamma(R,R',t,t',t'')=
    \frac{1}{\Omega^2}\sum_{k\leq k_F}    
    e^{-i\mathbf{k}\cdot\mathbf{R}} e^{-i\varepsilon_{k}t}
    \nonumber\\
    &\times\sum_{k'> k_F}
    e^{i\mathbf{k}'\cdot\mathbf{R}'}e^{-i\varepsilon_{k'}t'}
M(|\mathbf{k}'-\mathbf{k}|,\varepsilon_{k'}-\varepsilon_{k},t''),
\end{align}
$A_1$, $A_2$, $B_1$, and $B_2$ can be written as
\begin{align}
    A_1 &= \gamma(R',R'',t,0,t), \\
    A_2 &= \gamma(R'',R',t,0,0), \\
    B_1 &= \gamma(R',R'',0,t,0), \\
    B_2 &= \gamma(R'',R',0,t,-t).
\end{align}
The correlation hole is given by
\begin{align}
    \rho_\mathrm{c}(R,R',\theta;t<0) &=\frac{A_1+A_2}{G_0(R,t<0)}, 
    \nonumber\\
    \rho_\mathrm{c}(R,R',\theta;t>0) &=\frac{B_1+B_2}{G_0(R,t>0)} .
\end{align}


\end{document}